\documentclass{emulateapj}
\shorttitle{Early GRB Afterglows}
\shortauthors{Rykoff et al.}

\begin{document}

\newcommand\swift{\emph{Swift}}
\newcommand\betax{\beta_\mathrm{X}}
\newcommand\alphax{\alpha_\mathrm{X}}
\newcommand\alphao{\alpha_\mathrm{O}}
\newcommand\betaox{\beta_\mathrm{OX}}
\newcommand\fnuo{f_{\nu,\mathrm{O}}}
\newcommand\fnux{f_{\nu,\mathrm{X}}}
\newcommand\eiso{E_{\mathrm{iso}}}
\newcommand\eisoft{E_{\mathrm{iso},52}}
\newcommand\tpk{t_{\mathrm{pk}}}
\newcommand\epk{E_{\mathrm{pk}}}

\title{Looking into the Fireball: ROTSE-III and Swift Observations of Early
GRB Afterglows}

\author{
E.~S.~Rykoff,\altaffilmark{1},
F.~Aharonian\altaffilmark{2},
C.~W.~Akerlof\altaffilmark{3},
M.~C.~B.~Ashley\altaffilmark{4},
S.~D.~Barthelmy\altaffilmark{5},
H.~A.~Flewelling\altaffilmark{3},
N.~Gehrels\altaffilmark{5},
E.~G\"{o}\v{g}\"{u}\c{s}\altaffilmark{6},
T. G\"{u}ver\altaffilmark{7},
\"{U}.~K{\i}z{\i}lo\v{g}lu\altaffilmark{8},
H.~A.~Krimm\altaffilmark{5,9},
T.~A.~McKay\altaffilmark{3},
M.~\"{O}zel\altaffilmark{10},
A.~Phillips\altaffilmark{4},
R.~M.~Quimby\altaffilmark{11},
G.~Rowell\altaffilmark{12},
W.~Rujopakarn\altaffilmark{13},
B.~E.~Schaefer\altaffilmark{14},
D.~A.~Smith\altaffilmark{15},
W.~T.~Vestrand\altaffilmark{16},
J.~C.~Wheeler\altaffilmark{17},
J.~Wren\altaffilmark{16},
F.~Yuan\altaffilmark{3},
S.~A.~Yost\altaffilmark{18}
}

\altaffiltext{1}{TABASGO Fellow, Physics Department, University of California
  at Santa Barbara, 2233B Broida Hall, Santa Barbara, CA 93106, USA; erykoff@physics.ucsb.edu}
\altaffiltext{2}{Max-Planck-Institut f\"{u}r Kernphysik, Saupfercheckweg 1, 69117 Heidelberg, Germany}
\altaffiltext{3}{Department of Physics, University of Michigan, Ann Arbor, MI 48109, USA}
\altaffiltext{4}{School of Physics, Department of Astrophysics and Optics,
        University of New South Wales, Sydney, NSW 2052, Australia}
\altaffiltext{5}{NASA Goddard Space Flight Center, Laboratory for High Energy
        Astrophysics, Greenbelt, MD 20771, USA}
\altaffiltext{6}{Faculty of Engineering \& Sciences, Sabanc{\i} University,
  Orhanl{\i}-Tuzla 34956 {\.I}stanbul, Turkey}
\altaffiltext{7}{Department of Astronomy, University of Arizona, Tucson, AZ,
  85721, USA}
\altaffiltext{8}{Middle East Technical University, 06531 Ankara, Turkey}
\altaffiltext{9}{Universities Space Research Association, 10227 Wincopin
        Circle, Suite 212, Columbia, MD 21044, USA}
\altaffiltext{10}{\c{C}a\v{g} \"{U}niversitesi, Faculty of Arts and Sciences,
  Yenice-Tarsus/Mersin, Turkey}
\altaffiltext{11}{Division of Physics, Mathematics and Astronomy, California
  Institute of Technology, Pasadena, CA 91125, USA}
\altaffiltext{12}{School of Chemistry \& Physics, University of Adelaide, Adelaide 5005, Australia}
\altaffiltext{13}{Steward Observatory, University of Arizona, 933 North Cherry
  Avenue, Tucson, AZ 85721, USA}
\altaffiltext{14}{Department of Physics and Astronomy, Louisiana State
        University, Baton Rouge, LA 70803, USA}
\altaffiltext{15}{Guilford College, Greensboro, NC 27410, USA}
\altaffiltext{16}{Los Alamos National Laboratory, NIS-2 MS D436, Los Alamos,
NM 87545, USA}
\altaffiltext{17}{Department of Astronomy, University of Texas, Austin, TX
        78712, USA}
\altaffiltext{18}{Department of Physics, College of St. Benedict/St. John's
  University, Collegeville, MN 56321, USA}

\begin{abstract}

We report on a complete set of early optical afterglows of gamma-ray bursts
(GRBs) obtained with the ROTSE-III telescope network from March 2005 through
June 2007.  This set is comprised of 12 afterglows with early optical and
\swift/XRT observations, with a median ROTSE-III response time of
$45\,\mathrm{s}$ after the start of $\gamma$-ray emission (8~s after the GCN
notice time).  These afterglows span four orders of magnitude in optical
luminosity, and the contemporaneous X-ray detections allow multi-wavelength
spectral analysis.  Excluding X-ray flares, the broadband synchrotron spectra
show that the optical and X-ray emission originate in a common region,
consistent with predictions of the external forward shock in the fireball
model.  However, the fireball model is inadequate to predict the temporal decay
indices of the early afterglows, even after accounting for possible
long-duration continuous energy injection.  We find that the optical afterglow
is a clean tracer of the forward shock, and we use the peak time of the forward
shock to estimate the initial bulk Lorentz factor of the GRB outflow, and find
$100\lesssim\Gamma_0\lesssim1000$, consistent with expectations.

\end{abstract}


\section{Introduction}

The launch of the \emph{Swift Gamma-Ray Burst Explorer}~\citep{gcgmn04} has
brought considerable advancement to the study of gamma-ray bursts (GRBs).  The
rapid identification of GRBs by \swift{} Burst Alert
Telescope~\citep[BAT;][]{bbcfg05b}, combined with its excellent position
resolution, has allowed robotic automated ground based telescopes such as
ROTSE-III~\citep{akmrs03}, TAROT~\citep{kbag09}, RAPTOR~\citep{vbcfg04}, and
REM~\citep{zcgrt01} to respond promptly to GRBs with regularity, often taking
images contemporaneously with significant $\gamma$-ray emission.  Furthermore,
the \swift{} X-Ray Telescope~\citep[XRT;][]{hbnaa04} spectrometer provides soft
X-ray coverage of the tail of the prompt event and the early afterglow.
Combined, these observations provide an unprecedented view into the fireball of
the early GRB afterglow.

The ``fireball model'' of GRB emission (see \citet{p05} for a review) has been
successful in predicting the gross behavior of the late burst afterglow.
However, there are several inconsistencies between observations and modeling
for individual bursts, especially at the earliest times.  Most early X-ray
afterglows have a portion where the decay is significantly slower than
predicted by the fireball model, and this has been interpreted as evidence for
long-duration steady energy injection into the forward
shock~\citep[e.g.][]{nkgpg06}.  However, the cessation of early energy
injection would be expected to produce an achromatic light curve break, which
is not generally observed~\citep[e.g.][]{pmbng06}, although interpretation of
X-ray breaks is not always straightforward~\citep{rlbfs08}. Observations of
individual
bursts~\citep[e.g.][]{abbbb99,sraab03,rspaa04,wvwwe05,qryaa06,rmysa06,vwwag06,
rcccc06,mvmcd07,sdpvp07} have been interpreted in the context of various
models.  It is clear that for most early afterglows the ``closure
relationships'' \citep[e.g.][]{gs02}, which compare the spectral index of the
synchrotron emission to the temporal decay index, are often inconsistent
with the fireball model for individual bursts.

Early optical observations can also provide insight into the nature of the
fireball, especially by probing the onset of the afterglow, while the early
X-ray emission is often dominated by the tail end of the prompt burst emission.
However, very early optical detections have been difficult to obtain.  The
first optical flash observed contemporaneously from a GRB was from
GRB~990123~\citep{abbbb99}.  The $9^{th}$ magnitude optical flash was not
correlated with the high energy $\gamma$-ray emission, and its temporal
structure was consistent with reverse shock emission.  However, the optical
temporal sampling of the light curve was very limited, which made detailed
analysis impossible.  Other results from optical follow-up to BATSE bursts have
shown that these bright optical flashes are rare, but contemporaneous optical
detections of large numbers of bursts have not been possible until the
\swift{} era.  The connection between the prompt optical and $\gamma$-ray
emission is still not clear.  Although GRB~041219a~\citep{vwwfs05} and
GRB~050820a~\citep{vwwag06} appear to have a correlation between these two
components, this has not been seen for other bursts such as
GRB~050401~\citep{rykaa05}. \citet{yaaab07} have conducted a census of
ROTSE-III detections and deep non-detections of prompt optical counterparts,
and have not observed a strong correlation between the prompt optical and
$\gamma$-ray emission for these bursts.

Although connecting the high energy $\gamma$-ray emission with contemporaneous
optical emission can be instructive, this requires a large extrapolation, from
$10^{17}\,\mathrm{Hz}$ to $10^{14}\,\mathrm{Hz}$.  To study the full transition
from the prompt emission of the internal shock to the external forward shock we
require more complete wavelength coverage.  The XRT instrument is ideal
to help bridge the gap between the optical and $\gamma$-ray bands.  The X-ray
band is much less affected by Galactic and host absorption than UV or optical,
yet the soft X-rays provide a useful probe of the local equivalent hydrogen
column density.  Furthermore, the sensitivity of the XRT allows monitoring of
the high energy afterglow spectrum for tens of thousands of seconds.  However,
the early X-ray afterglow often contains flaring activity~\citep{bfcmr07} that
appear to originate in late internal
shocks~\citep[e.g.][]{lp07,kgmpb07,rlbfs08}.  Thus, when it is bright enough to
be detected at the early time, the optical emission might be more of a
``clean'' tracer of the external shock~\citep[e.g.][]{mvmcd07}.

The Robotic Optical Transient Search Experiment (ROTSE-III) array is a
worldwide network of 0.45~m robotic, automated telescopes, built for fast
($\sim 6\,\mathrm{s}$) responses to GRB triggers from satellites such as
\swift{}.  With four sites around the globe at Siding Spring Observatory,
Australia; McDonald Observatory, Texas; the H.E.S.S. site, Namibia; and the
Turkish National Observatory, Turkey, a ROTSE-III telescope is often ready for
a rapid response.  The ROTSE-III network commenced regularly responding to GRB
triggers from HETE-II~\citep{vvdjl99} in 2003~\citep{sraab03}.  After the
launch of \swift{} in late 2004, ROTSE-III began to respond to a significant
number of rapidly and well localized GRBs.  For $\sim30\%$ of \swift{} GRB
triggers, a ROTSE-III telescope is open in good weather and dark skies, and is
able to respond in under 1000~s.  We thus have a unique set of early GRB
afterglow light curves that are uniformly sampled.  We note that only a
fraction ($\sim 50\%$) of the bursts that are observed by ROTSE-III are
detected~\citep{yaaab07}, and thus ROTSE-III is only able to probe the brighter
afterglows.  When the broad-band open-filter ROTSE-III data are studied in
conjunction with early XRT observations, we can gain a deeper
understanding of the emission mechanisms of the early afterglow and its onset.

We have taken a complete set of 12 ROTSE-III afterglow light curves observed
between March 2005 and June 2007 for which we have contemporary XRT data.
These bursts are described in Section~\ref{sec:observations}.  These are a
complete census of ROTSE-III optical afterglows in this time period with early
($<500\,\mathrm{s}$) optical observations; XRT observations within $\sim
1000\,\mathrm{s}$; and more than three significant optical detections.  These
selection criteria excludes a few bursts with only marginal ROTSE-III
detections.  This collection of bursts have a median response time of
$45\,\mathrm{s}$ after the start of $\gamma$-ray emission ($8\,\mathrm{s}$
after the GCN notice time), providing a unique look at the earliest phases of
the optical afterglow.  For eight of these bursts, the ROTSE-III photometry is
being reported here for the first time; for the remainder, the afterglow data
have been published previously but have been re-analyzed here.

By studying these early afterglows as a set, we can discover the commonalities
as well as the differences.  Specifically, we can determine if the spectral and
temporal evolution of these afterglows is consistent with the fireball model
and a common emission mechanism.  For example, reverse shock emission has been
postulated as the source of the prompt optical flash of
GRB~990123~\citep{sp99b} and GRB~021004~\citep{fyktk03,kz03}, but has not been
observed in most early afterglows~\citep[e.g.][]{yaaab07,mmkgg08,kbag09}.  By
comparing the early optical and X-ray emission, we can also determine if the
shallow decay typically observed in X-ray afterglows is consistent with
continuous energy injection into the forward shock.  We can also study which
afterglow behaviors are part of a continuous distribution, and if any bursts
appear to be true outliers.  Finally, we can use the unprecedented early
optical coverage of many bursts to probe the onset of the afterglow.  This can
provide constraints on the bulk Lorentz factor of the outflowing
material~\citep[e.g.][]{mvmcd07}.  In similar vein, an analysis of 24 optical
afterglows detected within 10 minutes of the burst event was performed on GRB
detections from the Liverpool and Faulkes Telescopes~\citep{mmkgg08}.  They
find a wide range of early afterglow behavior, and several afterglows that
appear inconsistent with the fireball model.

In Section~\ref{sec:observations} we summarize the ROTSE-III observations used
in this paper.  Section~\ref{sec:reduction} describes the data reduction of the
ROTSE-III and \swift{} data.  In Section~\ref{sec:lightcurves} we present the
qualitative features of the multi-wavelength light curves for the 12 bursts.
Section~\ref{sec:spectra} compares the optical flux to the X-ray spectra for
the bursts.  Section~\ref{sec:temporal} describes quantitatively the temporal
evolution of these early afterglows, and Section~\ref{sec:risetimes} discusses
the large diversity of optical rise times in the context of the fireball
model.  Finally, we summarize our results and compare them to other recent work
in Section~\ref{sec:discussion}.

\section{Observations}
\label{sec:observations}

The ROTSE-III array is a worldwide network of 0.45~m robotic, automated
telescopes, built for fast ($\sim 6\,\mathrm{s}$) response to GRB triggers from
satellites such as \swift{}. They have wide ($1\fdg85 \times 1\fdg85$) fields
of view imaged onto Marconi $2048\times2048$ back-illuminated thinned CCDs, and
operate without filters.  The ROTSE-III systems are described in detail in
\citet{akmrs03}.

In this section, we describe the ROTSE-III observations of the bursts detailed
in this paper.  For several of these bursts (GRB~050525a, GRB~050922c,
GRB~060111b, GRB~060605, GRB~060729, GRB~060904b, GRB~061007, GRB~070611) the
full ROTSE-III photometry has not been previously published.  The remainder of
these bursts (GRB~050319, GRB~050401, GRB~050801, GRB~051109a) have been
published previously and are thus only described briefly.

\subsection{GRB 050525a}

On 2005 May 25, BAT detected GRB~050525a (\swift{} trigger 130088) at
00:02:53 UT.  The position was rapidly distributed as a GCN
notice~\citep{bcphb05}.  The burst had a $T_{90}$ duration of $8.8\,\mathrm{s}$
in the 15-350 keV band, and consisted of two peaks~\citep{cbbcf05,sbbcf08}.
The \swift{} satellite immediately slewed to the target, with the XRT beginning
observations 125~s after the start of the burst, and \swift/UVOT beginning
observations 65~s after the start of the burst.

ROTSE-IIIc and ROTSE-IIId both responded automatically to the GCN notice. The
first 5~s exposure from ROTSE-IIIc began at 00:08:56.7 UT, 8.7~s after the
notice, with bright ($>95\%$) lunar illumination.  ROTSE-IIIc took 10 5-s
exposures and 400 20-s exposures before morning twilight interrupted
observations.  The response by ROTSE-IIId was delayed approximately 30 minutes
due to bad weather.  ROTSE-IIId took 10 5-s images and 67 20-s images before
more bad weather interrupted observing.  Near real-time analysis of the
ROTSE-IIIc images detected a $15^{th}$ mag fading source at
$\alpha=18^h32^m32\fs6$ , $\delta=26\arcdeg20\arcmin23\farcs5$ (J2000.0) that
was not visible on the DSS red plates, which we reported via the GCN circular
e-mail exploder within 40 minutes of the burst~\citep{rys05}. The afterglow was
subsequently confirmed by UVOT~\citep{hbbrr05}.  Later spectroscopic
observations at Gemini-North determined the burst was at a redshift of
0.61~\citep{fcbp05}.

\subsection{GRB 050922c}

On 2005 September 22, BAT detected GRB~050922c (\swift{} trigger 156467)
at 19:55:50.4 UT.  The position was rapidly distributed as a GCN
notice~\citep{nbbcc05}.  The burst had a $T_{90}$ duration of $4.5\,\mathrm{s}$
in the 15-350 keV band, and consisted of a main peak with two
sub-peaks~\citep{kbbcc05,sbbcf08}.  The \swift{} satellite immediately slewed
to the target, with the XRT beginning observations 108~s after the trigger, and
UVOT beginning observations 111~s after the start of the burst.

ROTSE-IIId responded automatically to the GCN notice, beginning its first
exposure 6.8~s after the trigger time, at 19:58:42.8 UT, with moderately bright
($75\%$) lunar illumination.  The automated burst response included a set of
ten 5-s exposures, and 12 20-s exposures before bad weather interrupted
observing.  The weather cleared up briefly $\sim3000\,\mathrm{s}$ after the
burst, and 20 20-s exposures were taken before more bad weather set in
precluding further observations.  Near real-time analysis of the early
ROTSE-III images detected a bright, fading $15^{th}$ magnitude source at
$\alpha=21^h09^m33\fs1$, $\delta=-08\arcdeg45\arcmin29\farcs8$ (J2000.0) that
was not visible on the DSS red plates, which we reported via the GCN circular
e-mail exploder in less than 10 minutes from the initial BAT detection of the
burst~\citep{ryr05b}.  The afterglow was subsequently confirmed by
UVOT~\citep{nbbcc05}.  Later spectral follow-up determined the burst to
be at a redshift of $z=2.198$~\citep{jfptj05,pmmfc05,dpfca05}.

\subsection{GRB 060111b}

On 2006 January 11, BAT detected GRB~060111b (\swift{} trigger 176918)
at 20:15:43 UT.  The position was distributed as a Gamma-ray Burst Coordinates
Network (GCN) notice at 20:16:03 UT, during the prompt $\gamma$-ray
emission~\citep{pbbbc06}.  The burst had a $T_{90}$ duration of
$58.8\,\mathrm{s}$ in the 15-350 keV band, and consisted of two peaks
separated by about 55 seconds~\citep{tbbcc06,sbbcf08}.  The initial rise in
$\gamma$-ray emission began at 20:15:39 UT, and all times quoted in this paper
for this burst are relative to this obvious onset of the emission.  The
\swift{} satellite immediate slewed to the target, with XRT beginning
observations 79 s after the start of the burst, and UVOT beginning observations
84 seconds after the burst.  The initial \swift{} GCN circular at 20:53:53 UT
reported an X-ray counterpart and a UVOT ($B$-band) counterpart to
GRB~060111b at $\alpha=19^h05^m42\fs48$, $\delta=70\arcdeg22\arcmin33\farcs6$
(J2000.0).

ROTSE-IIId, at the Turkish National Observatory in Turkey, responded
automatically to the GCN notice, beginning its first exposure in less than 9~s,
at 20:16:12 UT, with bright ($>90\%$) lunar illumination.  The automated burst
response included a set of ten 5-s exposures and 408 20-s exposures before the
burst position dropped below our elevation limit.  Near real-time analysis of
the ROTSE-III images detected a rapidly fading $13^{th}$ magnitude source
coincident with the UVOT counterpart that was not visible on the Digitized Sky
Survey (DSS) red plates~\citep{yysa06}.  Due to the rapidly fading nature of
this counterpart and the bright lunar illumination, no spectral observations
were made to constrain the redshift of the burst.  We note that the \swift{}
trigger time was 4~s after the start of the GRB (see \S~\ref{sec:bat} for
details on the calculation of the burst start time).  In this paper, we
reference all times to the start of $\gamma$-ray emission at 20:15:39 UT.

\subsection{GRB 060605}

On 2006 June 05, BAT detected GRB~060605 (\swift{} trigger 213630) at
18:15:44 UT.  The position was distributed as a GCN notice at 18:16:28
UT~\citep{pbbgh06}.  The burst had a $T_{90}$ duration of $79.1\,\mathrm{s}$ in
the 15-350 keV band, and consisted of two overlapping FRED
peaks~\citep{sbbcf06,sbbcf08}.  The \swift{} satellite immediately slewed to
the target, with XRT beginning observations 93 s after the start of the burst,
and UVOT beginning observations 98 s after the burst.

ROTSE-IIIa, located at Siding Spring Observatory, Australia, responded
automatically to the GCN notice, beginning its first exposure in less than 6~s,
at 18:16:33.3 UT under good conditions.  The automated burst response included
a set of ten 5-s exposures, ten 20-s exposures, and 80 60-s exposures before
morning twilight interrupted observations.  Near real-time analysis of the
ROTSE-III images detected a variable $16^{th}$ magnitude source at
$\alpha=21^h28^m37\fs3$, $\delta=-06\arcdeg03\arcmin30\farcs6$ (J2000.0) that
was not visible on the DSS red plates, which we reported via the GCN circular
e-mail exploder less than 10 minutes from the initial BAT detection of the
burst~\citep{rs06,srsq06}.  The afterglow was subsequently confirmed by
UVOT~\citep{pbbgh06}.  Later spectral follow-up determined the burst to
be at a redshift of $z=3.78$~\citep{ps06,skrhl06,spfk07}.

\subsection{GRB 060729}

On 2006 July 29, BAT detected GRB~060729 (\swift{} trigger 221755) at
19:12:29 UT.  The position was distributed as a GCN notice at 19:13:26 UT
during the prompt $\gamma$-ray emission~\citep{gbbbc06}.  The burst had a
$T_{90}$ duration of 115.3~s in the 15-350 keV band, and consisted of the
initial $\gamma$-ray emission followed by two bright overlapping peaks at
$\sim80-90\,\mathrm{s}$~\citep{pbbcf06,sbbcf08}.  The \swift{} satellite
immediately slewed to the target, with XRT beginning observations 124~s after
the start of the burst, and UVOT beginning observations 135 s after the burst.
The initial \swift{} GCN circular at 19:53:56 UT reported an X-ray counterpart
and an optical (UVOT white filter) counterpart to GRB~060729 at
$\alpha=06^h21^m31\fs85$, $\delta=-62\arcdeg22\arcmin12\farcs7$ (J2000.0).

ROTSE-IIIa responded automatically to the GCN notice, beginning its first
exposure in less than 9~s, at 19:13:33.5 UT under good conditions.  The
automated burst response included a set of ten 5-s exposures, ten 20-s
exposures, and 35 60-s exposures before morning twilight interrupted
observations.  Near real-time analysis of the ROTSE-III images detected a
variable $16^{th}$ magnitude source coincident with the UVOT counterpart that
was not visible on the DSS red plates~\citep{qsrs06}.  Later spectral follow-up
determined the burst to be at a redshift of $z=0.54$~\citep{tljrg06}.  

\subsection{GRB 060904b}

On 2006 September 04, BAT detected GRB~060904b (\swift{} trigger 228006)
at 02:31:03 UT.  The position was rapidly distributed as a GCN notice at
02:31:17 UT during the prompt $\gamma$-ray emission~\citep{gbcch06}.  The burst
had a $T_{90}$ duration of $172\,\mathrm{s}$ in the 15-350 keV band, and
consisted of several peaks and long periods of
quiescence~\citep{mbbcf06,sbbcf08}.  The \swift{} satellite immediately slewed
to the target, with the XRT beginning observations 69~s after the start of the
burst, and UVOT beginning observations 70~s after the burst.

ROTSE-IIIc, at the H.E.S.S. site in Namibia, responded automatically to the GCN
notice, beginning its first exposure in less than 6~s, at 02:31:22.4 UT under
good conditions.  The automated burst response included a set of ten 5-s
exposures, 10 20-s exposures, and 85 60-s exposures before morning twilight
interrupted observations.  Near real-time analysis of the ROTSE-III images
detected a brightening $17^{th}$ magnitude source at $\alpha=03^h52^m50\fs5$,
$\delta=-00\arcdeg43\arcmin30\farcs9$ (J2000.0) that was not visible on the DSS
red plates, which we reported via the GCN circular e-mail exploder in less than
15 minutes from the initial BAT detection of the burst~\citep{rry06}. The
afterglow was subsequently confirmed by UVOT~\citep{gbcch06}.  Later
spectral follow-up determined the burst to be at a redshift of
$z=0.703$~\citep{fdmtc06}.

\subsection{GRB 061007}

On 2006 October 07, BAT detected GRB~061007 (\swift{} trigger 232683) at
10:08:08 UT.  The position was rapidly distributed as a GCN notice at
10:08:26~UT during the prompt $\gamma$-ray emission~\citep{scgpp06}.  The burst
had a $T_{90}$ duration of $75.3\,\mathrm{s}$ in the 15-350 keV band, and
consisted of three large peaks with a long faint tail~\citep{mbbcf06b,sbbcf08}.
The \swift{} satellite immediately slewed to the target, with the XRT beginning
observations 80~s after the start of the burst, and UVOT beginning observations
195~s after the burst.

ROTSE-IIIa responded automatically to the GCN notice, beginning its first
exposure in 9~s, at 10:08:35.2 UT, with bright ($>99\%$) lunar illumination.
The automated burst response included a set of ten 5-s exposures, and 460 20-s
exposures, before clouds interrupted observations.  Near real-time analysis of
the ROTSE-III images detected a bright, variable $13^{th}$ magnitude source at
$\alpha=03^h05^m19\fs6$, $\delta=-50\arcdeg30\arcmin02\farcs5$ (J2000.0) that
was not visible on the DSS red plates, which we reported via the GCN circular
e-mail exploder in less than 5 minutes from the initial BAT detection of the
burst~\citep{rr06}.  The afterglow was subsequently confirmed by
UVOT~\citep{scgpp06}.  Later spectral follow-up determined the burst to
be at a redshift of $z=1.26$~\citep{ocp06,jftr06}.

\subsection{GRB 070611}

On 2007 June 11, BAT detected GRB~070611 (\swift{} trigger 282003) at
01:57:13.9 UT.  The position was rapidly distributed as a GCN
notice~\citep{sbcgg07}.  The burst had a $T_{90}$ duration of
$12.2\,\mathrm{s}$ in the 15-350 keV band~\citep{sbbcf08}, with an additional
faint peak at $t_0+70\,\mathrm{s}$~\citep{bbcfg07}.  The \swift{} satellite
could not immediately slew to the target due to an Earth limb constraint, and
therefore the XRT and UVOT started observing the burst approximately 1 hour
after the trigger time.

ROTSE-IIIc responded automatically to the GCN notice, beginning its first
exposure 7.6~s after the trigger time, at 01:57:58.6 UT.  The automated burst
response included a set of ten 5-s exposures, ten 20-s exposures, and 130 60-s
exposures before morning twilight interrupted observations.  Initial analysis
of the ROTSE-III images did not reveal an optical counterpart~\citep{syry07}.
Later analysis after an XRT position was released~\citep{skp07} revealed a
late-rising optical counterpart at $\alpha=00^h07^m58\fs0$,
$\delta=-29\arcdeg45\arcmin19\farcs4$ (J2000.0) that was not visible on the DSS
red plates~\citep{ryy07}.  The afterglow was subsequently confirmed by
UVOT~\citep{slbgh07}.  Later spectral follow-up determined the burst to
be at a redshift of $z=2.04$~\citep{tjfmh07}.

\subsection{Other ROTSE-III Bursts}

Several other GRBs detected by ROTSE-III have been published previously and
modeled in detail.  Here we present a brief list of the other bursts described
in this paper.

\subsubsection{GRB~050319}

On 2005 March 14, BAT detected GRB~050319 (\swift{} trigger 111622).
ROTSE-IIIb had a rapid 8~s response to this burst, which is described in detail
in \citet{qryaa06}.  The burst began during a pre-planned slew of the
\swift{} satellite, and therefore the trigger time does not match the start of
$\gamma$-ray emission.  We follow \citet{qryaa06} in setting $t_0$ to
09:29:01.44 UT.

\subsubsection{GRB~050401}

On 2005 April 01, BAT detected GRB~050401 (\swift{} trigger 113130) at
14:20:15 UT.  ROTSE-IIIa had a rapid 6~s response to this burst, which is
described in detail in \citet{rykaa05}.  As the $\gamma$-ray emission preceded
the trigger time, we follow \citet{rykaa05} and refer all times for this burst
relative to 14:20:06 UT.

\subsubsection{GRB~050801}

On 2005 August 01, BAT detected GRB~050801 (\swift{} trigger 148522) at
18:28:02.1 UT.  ROTSE-IIIc had a rapid 8~s response to this burst, which is
described in detail in \citet{rmysa06}.  Although an optical spectrum was not
obtained for this afterglow, multi-wavelength optical and NUV observations with
UVOT have been used to calculate a photometric redshift of
$z\sim1.6$~\citep{dopbb07}.

\subsubsection{GRB~051109a}

On 2005 November 09, BAT detected GRB~051109a (\swift{} trigger 163136)
at 01:12:20 UT.  ROTSE-IIIb had a rapid 5~s response to this burst, which is
described in detail in \citet{ysraa07}.

\section{Data Reduction}
\label{sec:reduction}

In the interest of uniformity, we have used the same analysis for all of the
bursts presented in this paper.  The ROTSE-III optical data was processed using
the ROTSE photometry package RPHOT as described in \S~\ref{sec:rphot}.  The
analysis of the BAT and XRT observations are described in
\S~\ref{sec:bat} and \ref{sec:xrt}.  The analysis presented in this paper
is not intended to be a comprehensive study of the high energy emission of
these bursts.  More detailed spectral comparisons of simultaneous ROTSE-III and
BAT observations for all ROTSE-III bursts through GRB~061222 are
presented in \citet{yaaab07}.  

\subsection{ROTSE-III}
\label{sec:rphot}

The ROTSE-III images were bias-subtracted and flat-fielded by our automated
pipeline~\citep{rykoff05}.  We used SExtractor~\citep{ba96} to perform initial
object detection and to determine the centroid positions of the stars.  The
images are then processed with the RPHOT photometry program~\citep{qryaa06}
which performs relative photometry on magnitudes calculated with the DAOPHOT
PSF-fitting photometry package~\citep{stetson87}.  The unfiltered thinned
ROTSE-III CCDs have a peak response similar to an $R$-band filter.  The
magnitude zero point was calculated from the median offset of the fiducial
reference stars to the USNO~B1.0 $R$-band measurements to produce $C_R$
magnitudes.  When the signal-to-noise of individual images is too small for
detection, images are stacked in sets of 5, 10, or 20 to obtain deeper
exposures.  When a detection is not possible the $3\sigma$ upper limit is
quoted, as calculated from the local sky noise in a 1 FWHM aperture.  

The optical photometry and coincident X-ray flux measurements (see
\S~\ref{sec:xrt}) are listed in Table~\ref{tab:photom}.  We have converted the
ROTSE-III magnitudes to flux density ($f_\nu$) and flux by assuming the
unfiltered magnitudes are roughly equivalent to the $R_C$-band system, with
$\nu_{\mathrm{eff}} = 4.68\times10^{14}\,\mathrm{Hz}$~\citep[see,
e.g.,][]{rykaa05,rmysa06}.  When converting the photometric measurements
reported in Table~\ref{tab:photom} to flux and flux
density, we have corrected for Galactic absorption and extinction due to
Ly$\alpha$ absorption in the intergalactic medium (IGM).  To correct for
Galactic extinction we used the values of $A_R$ from \citet{sfd98}, which are
reported in Table~\ref{tab:extabs}.  For the bursts at a redshift of $z\gtrsim
2.0$, the Ly$\alpha$ absorption cuts into the ROTSE-III bandpass.  To correct
for this, we follow the method outlined in \citet{rstmf07}.  We first assume
the spectral energy distribution of the optical afterglow has a power-law form
$f_\nu(\nu) \propto \nu^{\beta}$, with $\beta = -0.75$.  This spectrum is
folded with the Ly$\alpha$ absorption in the IGM using the model of
\citet{m05}, and then with the ROTSE spectral response.  The fraction of flux
lost to absorption by the IGM is converted to an equivalent magnitude offset,
reported in Table~\ref{tab:extabs}.  Note that this value is not very sensitive
to the assumption of the input spectrum: changing $\beta$ by $\pm0.25$ changes
the equivalent magnitude offset by $\lesssim 0.05$.

\begin{deluxetable}{cccc}
\tablecaption{Galactic Extinction and IGM Absorption\label{tab:extabs}}
\tablewidth{0pt}
\tablehead{
\colhead{GRB} &
\colhead{$z$} &
\colhead{$\Delta m_{IGM}$} &
\colhead{$A_R$}
}
\startdata
GRB~050319 & 3.24 & 0.147 & 0.029\\
GRB~050401 & 2.90 & 0.07 & 0.174\\
GRB~050525a & 0.61 & 0.0 & 0.254\\
GRB~050801 & 1.6   & 0.0 & 0.257\\
GRB~050922c & 2.20 & 0.01 & 0.276\\
GRB~051109a & 2.35 & 0.01 & 0.508\\
GRB~060111b & 1.0? & 0.0 & 0.297\\
GRB~060605 & 3.80 & 0.38 & 0.132\\
GRB~060729 & 0.54 & 0.0 & 0.145\\
GRB~060904b & 0.70 & 0.0 & 0.463\\
GRB~061007 & 1.26 & 0.0 & 0.055\\
GRB~070611 & 2.04 & 0.01 & 0.033\\
\enddata
\end{deluxetable}

\subsection{\swift/BAT}
\label{sec:bat}

The BAT and XRT observations were processed using the packages and tools
available in HEASOFT version 6.1\footnote{See
http://heasarc.gsfc.nasa.gov/docs/software/lheasoft}.  Initial mask-weighting
on the raw event files was performed with {\tt batmaskwtevt} using standard
quality cuts.  Light curves were generated with a fixed signal-to-noise ratio
(SNR) of 6.0 in the 15-150 keV energy band with {\tt batbinevt}.  To obtain
spectral files, we follow the standard BAT analysis from the BAT
DIGEST\footnote{http://swift.gsfc.nasa.gov/docs/swift/analysis/bat\_digest.html}.
The tool {\tt batbinevt} was used to extract a spectral ({\tt pha}) file with
the standard 80 channels over the desired time range (see below).  The tool
{\tt batupdatephakw} was used to update the BAT ray tracing columns in the
spectral file to correct for spacecraft slews during the burst.  The tool {\tt
batphasyserr} was used to calculate the systematic error, and finally {\tt
batdrmgen} was used to generate a spectral response ({\tt rsp}) file.

For most bursts, we use the BAT trigger time as the start time of the burst
($t_0$).  This is the start of the time interval in which a rate increase was
first seen on board the \swift{} satellite.  For three bursts (GRB~050319,
GRB~050401, and GRB~060111b; see \S~\ref{sec:observations}) the $\gamma$-ray
emission is significantly detected prior to $t_0$, and we have adjusted $t_0$
accordingly.  Using the light curves generated with {\tt batbinevt} we
have confirmed that the quoted values of $t_0$ can be equivalently defined as
the time at which the $\gamma$-ray flux was detected with SNR$>6.0$, with a
typical error of $\pm5\,\mathrm{s}$.

For each burst we calculate the time-averaged spectrum using XSPEC version
11.3.2~\citep{a96}.  Many of the bursts exhibit significant spectral evolution,
generally from hard-to-soft, as is seen for most GRBs~\citep[e.g.][]{p05}.  For
the purposes of this work, however, it is simpler to use the time-averaged
spectrum over the duration of the burst to obtain a straightforward conversion
from count-rate to flux in the 15-150 keV band.  Detailed comparisons of
simultaneous BAT and ROTSE-III detections of these bursts are described in
\citet{yaaab07}.  Each of the BAT spectra was well-fit by a simple power-law
except for GRB~050525a, which was fit by a GRB model~\citep{bmfsp93}.  The
resulting spectral indices are shown in Table~\ref{tab:batspec}.  To display
the BAT light curves on the same plots as the XRT light curves, we have
extrapolated the BAT spectra to the XRT range (0.3-10 keV) using the average of
the time-averaged BAT power-law index and the XRT power-law index.

\begin{deluxetable}{ccc}
\tablecaption{BAT Spectral Indices\label{tab:batspec}}
\tablewidth{0pt}
\tablehead{
\colhead{GRB} &
\colhead{Fit Time Range (s)} &
\colhead{$\Gamma$}
}
\startdata
GRB~050319 & $0 - 170$ & $2.09\pm0.20$\\
GRB~050401 & $0 - 50$ & $1.48\pm0.08$\\
GRB~050525a & $-10 - 20$ & $0.97\pm0.15$\tablenotemark{a}\\
GRB~050801 & $-10 - 50$ & $2.06\pm0.20$\\
GRB~050922c & $-2 - 4$ & $1.33\pm0.05$\\
GRB~051109a & $-10 - 50$ & $1.51\pm0.25$\\
GRB~060111b & $-10 - 100$ & $0.90\pm0.18$\\
GRB~060605 & $0 - 25$ & $1.37\pm0.19$\\
GRB~060729 & $-5 - 150$ & $1.82\pm0.15$\\
GRB~060904b & $-10 - 230$ & $1.72\pm0.16$\\
GRB~061007 & $-10 - 70$ & $1.04\pm0.03$\\
           & $70 - 300$ & $1.75\pm0.10$\\
GRB~070611 & $-3 - 10$ & $1.61\pm0.27$\\
\enddata
\tablenotetext{a}{The spectrum of GRB~050525a is well-fit by a GRB model
  function, with $\epk = 79\pm15\,\mathrm{keV}$ and the high energy
  index fixed at $-2.5$.  The quoted value of $\Gamma$ is the equivalent low
  energy index.}
\end{deluxetable}

For one burst, GRB~061007, we performed a slightly different analysis for a
better comparison of the BAT light curve to the early XRT light curve.  For
this burst, the main event was significantly harder ($\Gamma \sim 1.0$) than
the long tail that was detected coincident with the X-ray afterglow ($\Gamma
\sim 1.8$).  Thus, we have split the spectral analysis into two time bins, from
$T-T_{\mathrm{trig}}<70\,\mathrm{s}$ and
$70\,\mathrm{s}<T-T_{\mathrm{trig}}<300\,\mathrm{s}$.  This shows a better
representation of the connection of the prompt event to the early afterglow.
Most of the other bursts did not display such dramatic evolution in their
spectral indices.  The other exception is GRB~060111b, although since there are
no multi-wavelength observations contemporaneous with the first peak we did not
see the need for a special correction.

\subsection{\swift/XRT}
\label{sec:xrt}

The XRT observations were processed with a pipeline that is described in
\citet{rmst07}.  Initial event cleaning was performed with {\tt xrtpipeline}
using standard quality cuts, using event grades 0--2 in WT mode (0--12 in PC
mode).  For the WT mode data, source extraction was performed with {\tt
  xselect} in a rectangular box 20 pixels wide and 40 pixels long.  Background
extraction was performed with a box 20 pixels wide and 40 pixels long far from
the source region.  For the PC mode data, source extraction was performed with
a 30 pixel radius circular aperture, and background extraction was performed
with an annulus with an inner (outer) radius of 50 (100) pixels.

After event selection, exposure maps were generated with {\tt xrtexpomap} and
ancillary response function ({\tt arf}) files with {\tt xrtmkarf}.  The latest
response files (v008) were used from the CALDB database.  All spectra
considered in this paper were grouped to require at least 20 counts per bin
using the ftool {\tt grppha} to ensure valid results using $\chi^2$ statistical
analysis.  Spectral fits were made with XSPEC in the 0.3--10 keV range.  All of
the X-ray flux measurements, unless otherwise noted, are in the 0.3--10 keV
range. The uncertainties reported in this work are 90\% confidence errors,
obtained by allowing all fit parameters to vary simultaneously.
 
Several of the PC observations were slightly affected by pile-up, especially in
the early observations.  When observations suffer from pile-up, multiple soft
photons can be observed at nearly the same time, and appear as a single hard
photon.  Pile-up correction was performed using spectral fitting, following the
method described in \citet{rcccc06} and \citet{rmst07}.

For the purpose of generating light curves, we have calculated the
time-averaged XRT spectra to obtain a conversion from count-rate in the 0.3-10
keV band to unabsorbed flux in the 0.3-10 keV band.  For each of the bursts
except for GRB~060927 (discussed below) we have fit the spectrum with an
absorbed power-law, using the {\tt wabs} absorption model~\citep{mm83}.  When
fitting combined XRT data sets (e.g. WT mode spectra; pile-up corrected PC
spectra; and non piled-up PC spectra) we tie the equivalent hydrogen column
density ($n_H$) and photon index across each data set, and allow the
normalizations to float between data sets, as the X-ray afterglow varies with
time.  To generate X-ray light curves, we bin the source events with a fixed 50
counts per time bin before background subtraction.  This ensures a roughly
equal signal-to-noise across the duration of the observation.

The X-ray light curve of GRB~060729 was very bright and the spectral shape was
varying quite rapidly.  During the WT observations from 150~s--356~s
post-burst, the spectrum is better fit by a soft GRB model function ($\epk =
2.1\pm0.5\,\mathrm{keV}$) with absorption fixed to the Galactic
value~\citep{dl90} than by an absorbed power-law, which would require
absorption that is correlated with the intensity.  The merits of this fitting
function are discussed in detail in \citet{bk07}.

We also use the X-ray spectra to estimate $n_H^z$, the equivalent hydrogen
column density at the redshift of the burst.  We first fit each XRT spectrum
with an absorbed power-law, to estimate the total equivalent hydrogen column
density, $n_H^T$.  If $n_H^T$ is significantly greater (at $>90\%$ confidence)
than the Galactic $n_H^G$ at the position of the burst~\citep{dl90}, then we
consider the afterglow to have a significant excess in $n_H$.  We then re-fit
the spectrum with a new absorption component at the redshift of the GRB, while
fixing $n_H^G$ at the Galactic value.  

The X-ray spectral indices and $n_H$ values for the 12 bursts in this paper are
listed in Table~\ref{tab:betax}.  The spectral index, $\betax$ is defined as
$\betax = 1-\Gamma$, for the power-law model $f_\nu \propto \nu^\betax$.  With
the exception of the aforementioned GRB~060729, we have confirmed that none of
the spectral indices varied significantly across light curve breaks calculated
in \S~\ref{sec:plawfits} and detailed in Table~\ref{tab:xrtplaws}.  The
observed lack of spectral evolution is consistent with a much more detailed
review of XRT light curves~\citep{rlbfs08}.  We note that all of the spectral
indices cluster around $\betax \sim -1.0$, which we discuss in greater detail
below.

\begin{deluxetable*}{ccccccc}
\tablecaption{X-ray Spectral Indices and $n_H$\label{tab:betax}}
\tablewidth{0pt}
\tabletypesize{\scriptsize}
\tablehead{
\colhead{GRB} &
\colhead{$z$} &
\colhead{Fit Time Range (s)} &
\colhead{$\betax$} &
\colhead{$n_H^G$\tablenotemark{a}} & 
\colhead{$n_H^T$\tablenotemark{b}} &
\colhead{$n_H^z$\tablenotemark{c}}\\
& & (s) & & $(10^{22}\,\mathrm{cm}^{-2})$ & $(10^{22}\,\mathrm{cm}^{-2})$ & $(10^{22}\,\mathrm{cm}^{-2})$
}
\startdata
GRB~050319 & 3.24 & $219-13362$ & $-0.99\pm0.16$ & $0.011$ & $<0.039$ & \\
GRB~050401 & 2.9 & $133-10000$ & $-0.99\pm0.05$ & $0.049$ & $0.137\pm0.013$ & $1.4\pm0.2$\\
GRB~050525a & 0.606 & $128-87212$ & $-0.98\pm0.06$ & $0.091$ & $0.18\pm0.05$ & $0.20\pm0.10$\\
GRB~050801 & 1.6 & $59-56602$ & $-0.9\pm0.2$ & $0.07$ & $0.06\pm0.03$ & \\
GRB~050922c & 2.2 & $107-69769$ & $-1.09^{+0.07}_{-0.04}$ & $0.057$ &
$0.070\pm0.015$ & \\
GRB~051109a & 2.35 & $119-26874$ & $-1.10\pm0.10$ & $0.174$ & $0.26\pm0.03$ & $1.0\pm0.4$\\
GRB~060111b & 1.0 & $83-70485$ & $-1.24\pm0.16$ & $0.069$ & $0.28\pm0.07$ & $0.8\pm0.3$\\
GRB~060605 & 3.80 & $88-74165$ & $-1.00\pm0.07$ & $0.051$ & $0.07\pm0.04$ & \\
GRB~060729 & 0.54 & $130-356$ & $-1.0\pm0.2$\tablenotemark{d} & $0.049$ & &\\
           &      & $356-12256$ & $-1.17\pm0.06$ & & $0.11\pm0.02$ & $0.11\pm0.04$\\
GRB~060904b & 0.70 & $70-40797$ & $-1.13\pm0.04$ & $0.111$ & $0.30\pm0.05$ & $0.49\pm0.15$ \\
GRB~061007 & 1.26 & $82-25372$ & $-0.97\pm0.02$ & $0.021$ & $0.19\pm0.03$ & $0.67\pm0.13$\\
GRB~070611 & 2.04 & $3288-45631$ & $-0.8^{+0.4}_{-0.5}$ & $0.013$ & $<0.15$ & \\
\enddata
\tablenotetext{a}{$n_H^G$ is the Galactic equivalent hydrogen column density from
  \citet{dl90}}
\tablenotetext{b}{$n_H^T$ is the total equivalent hydrogen column density,
  assuming all the absorption is at $z=0$.}
\tablenotetext{c}{$n_H^z$ is the host equivalent hydrogen column density, after
  fixing the $n_H (z=0) = n_H^G$.}
\tablenotetext{d}{$\betax$ is the low-energy component of a GRB model function,
as described in \S~\ref{sec:xrt}}
\end{deluxetable*}

\section{Multi-Wavelength Light Curves}
\label{sec:lightcurves}

We have assembled multi-wavelength BAT, XRT, and ROTSE-III light curves for the
12 bursts described in \S~\ref{sec:observations}.  The BAT analysis is
described in \S~\ref{sec:bat}, the XRT analysis is described in
\S~\ref{sec:xrt}, and the ROTSE-III analysis is described in
\S~\ref{sec:rphot}.  The BAT fluxes, calculated in the 15-150 keV range, have
been extrapolated to the XRT 0.3-10 keV band as described above, for an easier
comparison.  For plotting purposes, all times have been scaled by $(1+z)$ to
account for cosmological time dilation.  Of this set of afterglows, GRB~060111b
is the only burst without a redshift estimate.  As described in
\S~\ref{sec:observations}, the detection of this burst in the UV constrains the
redshift to be $\lesssim 1.5$.  We have therefore assumed a fiducial redshift
of 1.0 for GRB~060111b.

\begin{figure*}
\plotone{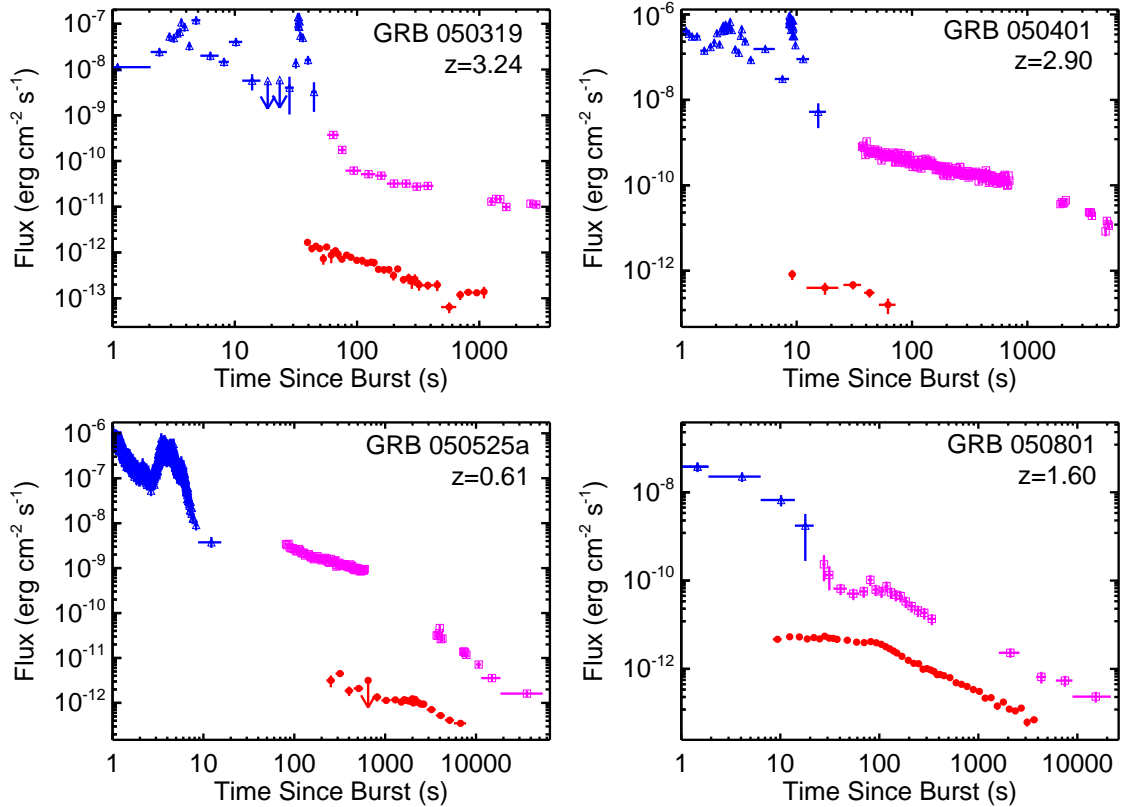}
\caption{\label{fig:combo1} Multi-wavelength light curves for four bursts.  The
  BAT data (blue triangles) has been extrapolated to the X-ray regime as described
  in \S~\ref{sec:bat}.  The XRT data is shown with magenta squares, and the ROTSE-III
  data with red circles.  In all cases the optical flux is below the X-ray
  flux.  The time axis has been corrected for cosmological time
  dilation.  \emph{GRB~050319:} The optical data does not show a deviation from
  a simple power-law, while the X-ray data shows the typical steep-flat
  evolution. \emph{GRB~050401:} Neither the optical nor the X-ray data shows
  deviations from simple power-laws. \emph{GRB~050525a:} The optical light
  curve shows a steepening at $\sim2000\,\mathrm{s}$, while the X-ray light
  curve shows a slightly more complicated evolution. \emph{GRB~050801:} The
  optical light curve shows a steepening at $\sim100\,\mathrm{s}$, and the
  X-ray light curve shows a very similar morphology.}
\end{figure*}

\begin{figure*}
\plotone{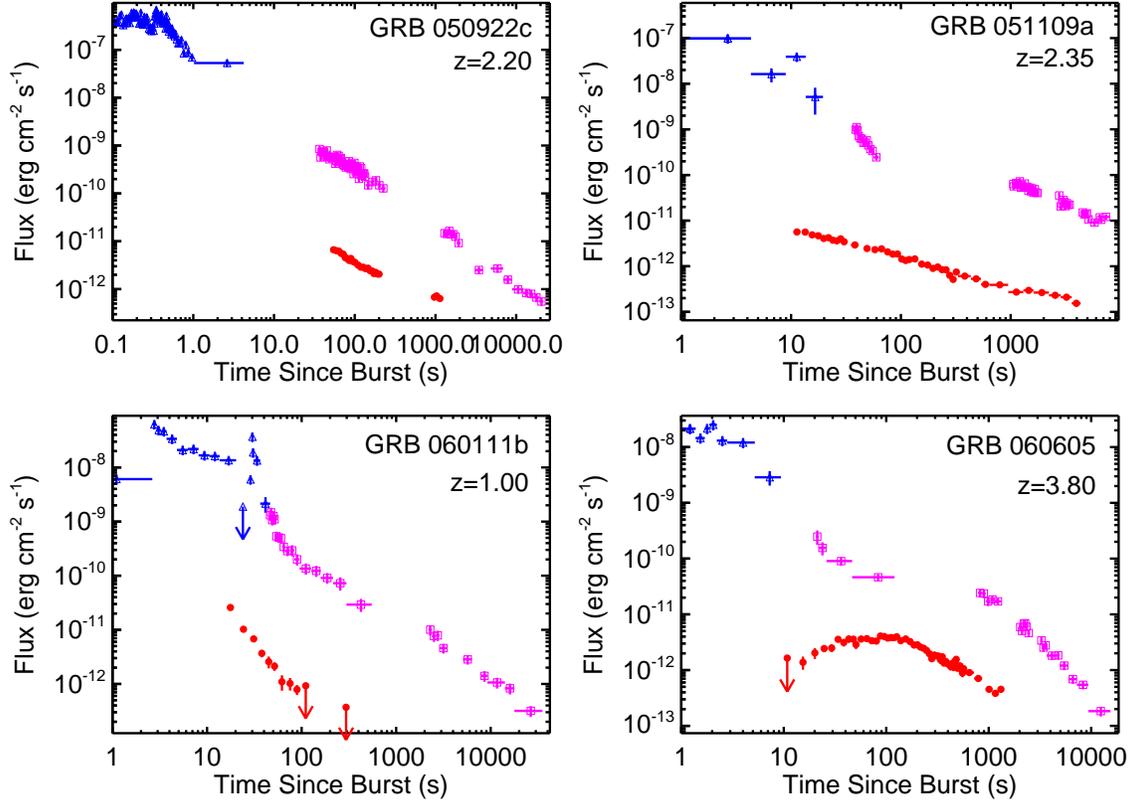}
\caption{\label{fig:combo2} Multi-wavelength light curves for four bursts, with
  the same symbols as Figure~\ref{fig:combo1}.  The time axis has been
  corrected for cosmological time dilation.  \emph{GRB~050922c:} Both the
  optical and X-ray light curves show a similar morphology, with a simple
  power-law decline. \emph{GRB~051109a:} The optical light curve follows a
  simple power-law, while the X-ray light curve shows the canonical
  steep-shallow-steep morphology, although most of the shallow section needs to
  be inferred from an interpolation over the orbital gap. \emph{GRB~060111b:}
  The rapidly decaying optical light curve peaks before the second $\gamma$-ray
  peak at $\sim30\,\mathrm{s}$. Note that the time axis has been scaled to an
  approximate redshift of $z=1.0$. \emph{GRB~060605:} The optical light curve
  shows a slow rise and decay, peaking at $\sim100\,\mathrm{s}$, while the
  contemporaneous X-ray light curve shows the typical steep-shallow-steep
  canonical form.}
\end{figure*}

\begin{figure*}
\plotone{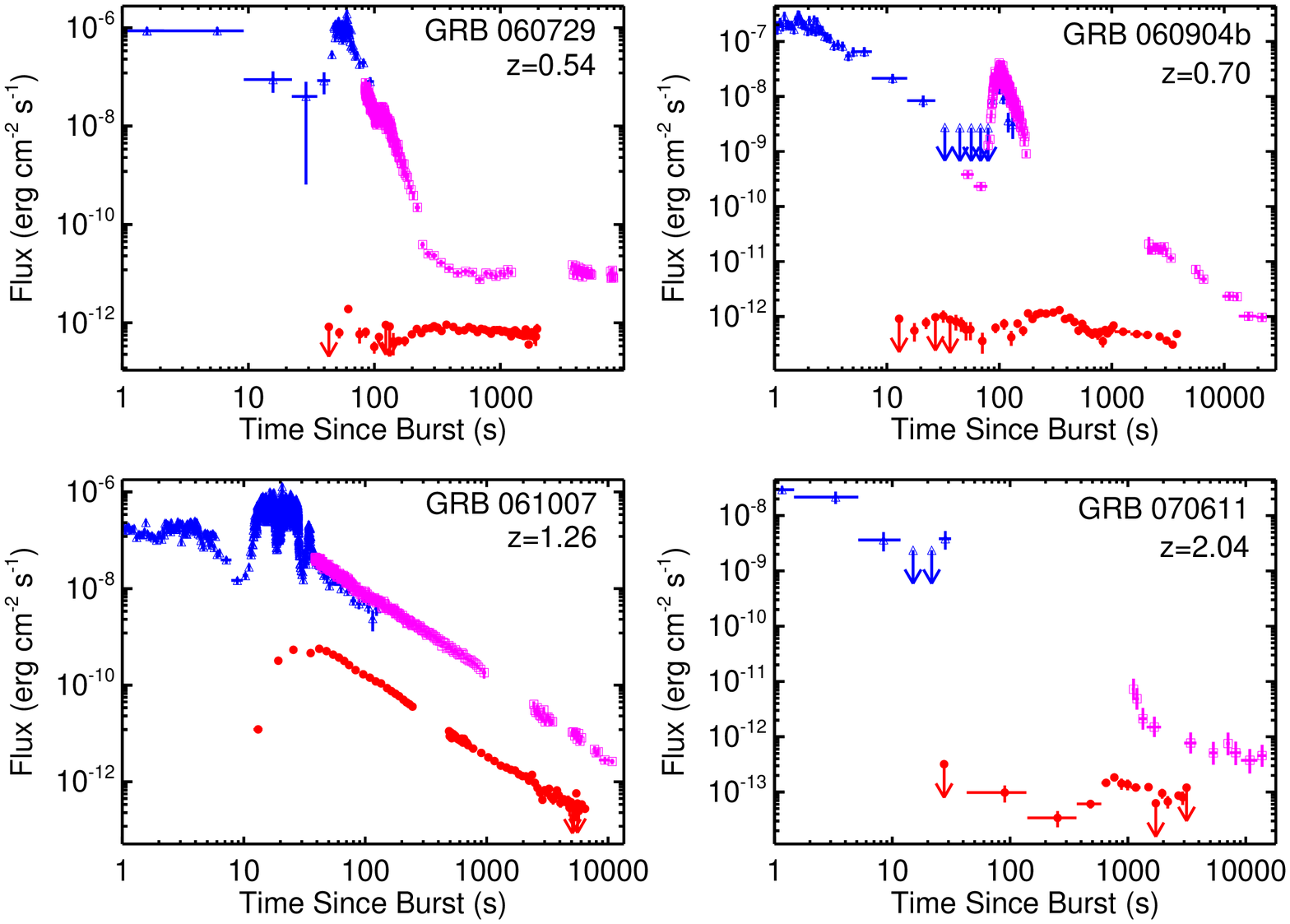}
\caption{\label{fig:combo3} Multi-wavelength light curves for four bursts, with
  the same symbols as Figure~\ref{fig:combo1}.  The time axis has been
  corrected for cosmological time dilation.  \emph{GRB~060729:} The optical
  light curve shows two peaks, the first an early flare at $\sim60\,\mathrm{s}$
  perhaps coincident with
  one of the $\gamma$-ray peaks, and the second around $300-500\,\mathrm{s}$
  (see \S~\ref{sec:calcpk}). The X-ray light curve shows a typical
  steep-shallow decay, with a flare around $\sim100\,\mathrm{s}$ that is not
  apparent in the contemporaneous optical light curve.  After
  $\sim300\,\mathrm{s}$, both the X-ray and optical decays are exceptionally
  shallow, resulting in a very long-lived X-ray
  afterglow~\citep{ggwrc07,gbxwz09}. \emph{GRB~060904b:} The optical light
  curve is complex for this burst, with short-term variability and an apparent
  peak at $\sim30\,\mathrm{s}$, followed by a smoother evolution with a peak at
  $\sim300\,\mathrm{s}$.  The X-ray light curve shows a giant flare at
  $\sim100\,\mathrm{s}$ that is not apparent in the contemporaneous optical
  light curve. \emph{GRB~061007:} The optical light curve shows a dramatic
  rise, brightening by over a factor of 50 in less than 5~s, followed by two
  peaks and a steady power-law decline.  The $\gamma$-ray light curve shows
  multiple peaks that are not contemporaneous with the optical peaks, followed
  by a steady decline in the X-rays that tracks the optical
  decline. \emph{GRB~070611:} There are hints of an early, faint optical peak
  around $\sim100\,\mathrm{s}$, and a pronounced peak at $\sim700\,\mathrm{s}$.
  The X-ray light curve is not well sampled due to an orbital break, but there
  is the hint of the tail of a flare around $\sim1000\,\mathrm{s}$, followed by
  a shallow decay. }
\end{figure*}

The multi-wavelength light curves are shown in
Figures~\ref{fig:combo1}--\ref{fig:combo3}.  The BAT flux values (extrapolated
to 0.3-10 keV) are blue triangles; the XRT flux values (0.3-10 keV) are magenta
squares, and the ROTSE-III flux values are the red circles.  In all cases, the
optical flux is below the X-ray flux.

\subsection{Qualitative Comparisons}

The temporal behavior of the earlier afterglows is, at first glance, quite
diverse.  For several of the bursts -- GRB~050319, GRB~050401, GRB~050525a,
GRB~050922c, GRB~051109a, and GRB~060111b -- the optical afterglow is already
fading by the time of the first ROTSE-III exposure.  For some bursts, this is
as soon as 10~s after the start of $\gamma$-ray emission.  Other afterglows are
seen to rise more slowly.  The optical afterglows of GRB~060605, GRB~060729,
GRB~060904b, and GRB~070611 peak several hundred seconds after the start of
$\gamma$-ray emission.  The afterglow of GRB~061007 shows the
most dramatic rise, brightening by over a factor of 50 in the optical in less
than 5~s.  The diversity of rise times and a possible physical origin are
discussed in \S~\ref{sec:risetimes}.

After the initial optical rise, if it is observed, the optical afterglow
typically fades as a power-law, although substructure is seen in some bursts.
The X-ray afterglow usually follows the ``canonical''
shape~\citep[e.g.][]{nkgpg06,rlbfs08}.  This consists of a steep initial
decline, a shallow plateau, and another power-law decline.  Often there are
X-ray flares superimposed on the canonical afterglow shape, which we observe
for GRB~060729 and GRB~060904b.  The steep initial decline of the X-ray
afterglow has been interpreted as the tail of the prompt emission, possibly
caused by high-latitude burst emission or the curvature
effect~\citep{kp00b,lzowa06,zlz07}.  This interpretation is supported by the
fact that the steep early X-ray decline links up with the tail end of the
$\gamma$-ray emission.  \citet{kgmpb07} have also pointed out that for
some bursts, later peaks detected by BAT have the same spectral and temporal
properties as X-ray flares.  Therefore, it may be completely arbitrary to
distinguish between the steep initial decline of the X-ray afterglow and an
X-ray flare.

At the earliest times, the optical afterglows do not show the same steep
decline as the X-ray emission.  This suggests that the optical and X-ray
emission originate from different regions at the start of the burst: the X-ray
emission is dominated by the internal shock emission which produced the GRB
itself, and the optical emission is dominated by the onset of the forward
external shock.  Similarly, we do not observe optical flares contemporaneously
with the X-ray flares.  This is consistent with the interpretation of the X-ray
flares as late internal shock
emission~\citep[e.g.][]{bfcmr07,bk07,lp07,kgmpb07}.  If, instead, the X-ray
flares were caused by density changes in the external medium, one would expect
a similar brightening in the optical afterglow which we do not observe.

The shallow plateau that is usually observed in the early X-ray afterglow is
significantly less steep than predicted in the standard fireball model.
Therefore, it has been interpreted as evidence for long-duration energy
injection into the external forward shock~\citep[e.g.][]{nkgpg06}.  If this
hypothesis were correct, we would expect that (a) the decay rate of the
contemporaneous optical afterglow is significantly less steep than predicted in
the basic fireball model and (b) at the cessation of the energy injection
episode there will be an achromatic break observed in both optical and X-ray
wavelengths.  For the ROTSE-III afterglows, we typically observe that the
shallow X-ray decay is accompanied by a shallow optical decay, but this is not
always the case, as discussed in \S~\ref{sec:temporal}.  Additionally, we do
not typically observe an achromatic break at the end of the shallow decay
phase.  This has also been noted by \citet{pmbng06}, who showed that the break
times associated with the end of continuous energy injection are usually not
consistent between the optical and X-ray afterglows.  However, the limited
temporal sampling of the afterglows described in this paper makes these
comparisons challenging.

After the initial optical rise and/or rapid X-ray decay, and excluding X-ray
flares, for all the afterglows we observe that the optical and X-ray afterglows
display similar trends.  As discussed in detail in \S~\ref{sec:spectra}, the
afterglows that are brighter in optical also tend to be brighter in X-rays.
And, as discussed in \S~\ref{sec:temporal}, the afterglows that fade rapidly in
the optical also fade rapidly in the X-rays (e.g. GRB~061007) while the
afterglows that fade slowly in the optical also fade slowly in X-rays
(e.g. GRB~060729).

\section{Broadband Spectra}
\label{sec:spectra}

\subsection{Fireball Model}

In the fireball model, the afterglow is produced by synchrotron emission from
shock-accelerated electrons.  If the optical and X-ray emission are from the
same emission region, there should be a relationship between the spectral and
temporal evolution of the optical and X-ray flux density.  This relationship
depends on the shape of the synchrotron spectrum, especially the location of
the various break frequencies.  Note that we have adopted the convention that
the flux density can be described as a local power law in both time and
frequency, such that $f_\nu \propto t^\alpha \nu^\beta$.  Here, $f_\nu$ is the
flux density in units of
$\mathrm{ergs}\,\mathrm{cm}^{-2}\,\mathrm{s}^{-1}\,\mathrm{Hz}^{-1}$, $\alpha$
is the temporal power-law index, and $\beta$ is the spectral power-law index.
Due to the fact that we observe afterglows both rising and fading, our
conventions differ from some other authors in that we explicitly quote the sign
of the power-law indices.

\citet{gs02} have compiled a useful list of the various spectral relationships
that may be observed during the self-similar evolution of the GRB afterglow in
the fireball model.  In the more commonly observed ``slow cooling'' regime,
where the cooling frequency, $\nu_\mathrm{c}$ is above the peak synchrotron
frequency, $\nu_\mathrm{m}$, then the flux density above the cooling frequency
is given by $f_\nu \propto \nu^{-p/2}$, where $p$ is the spectral index of the
input electron energies, such that $N(\gamma) \propto \gamma^{-p}$.  Below the
cooling frequency, $f_\nu \propto \nu^{(1-p)/2}$.  It is expected that the
X-ray band will always be above $\nu_c$ and the optical band will be above or
below $\nu_c$ depending on the microphysical parameters in the shock, as well
as the time elapsed from the start of the burst.  For an external shock
expanding into a constant density medium, this implies that we may expect a
light curve break when $\nu_c$ passes through the optical waveband.  We note
that for the afterglows in this paper, virtually all of the X-ray spectra have
a spectral index consistent with $\betax \sim -1.0$.  This corresponds to an
electron spectral index $p\sim2$, which we have taken as our fiducial value for
all the bursts in this paper.  The implied electron spectral index is different
if we use the temporal decay index $\alpha$ to estimate $p$, as is discussed in
\S~\ref{sec:plawcompare}.

If the optical and X-ray emission both originate in the forward external shock,
then we can extrapolate the X-ray synchrotron spectrum to the ROTSE-III
bandpass to predict the optical flux density.  For simplicity, we first neglect
optical extinction due to dust in the host galaxy.  The maximum optical flux
density, $\fnuo$, corresponding to a given X-ray flux density, $\fnux$, will
occur when the two bands are in the same synchrotron regime (e.g., the cooling
break $\nu_c$ is below the optical band).  Given the X-ray spectral index of
$\betax\sim-1.0$, the broadband spectral index between the optical and X-ray
band should also be $\betaox \sim -1.0$.  The minimum $\fnuo$ corresponding to
a given $\fnux$ will occur when the cooling break $\nu_c$ is just below the
X-ray band, which yields a relatively flat broadband spectral index, $\betaox
\sim -0.5$.

On the other hand, if the optical and X-ray emission are not from the same
region, then we do not expect to observe this simple relationship.  For
example, if the X-ray flares or steep X-ray decline are from internal shocks
caused by late activity of the central engine, then they should not be part of
the same synchrotron spectrum as the optical afterglow.  Similarly, there may
be an optical flash caused by reverse shock
emission~\citep{mr97,sp99,sp99b,k00}, which would instead peak in the
NIR/optical/NUV wavelengths.  Thus, we expect this optical flash may be
\emph{overluminous} compared to an extrapolation of the X-ray emission.

\subsection{Optical and X-ray Comparison}

We have compared the optical and X-ray flux density ($\fnuo$ and $\fnux$) at
multiple epochs for each of the bursts in this paper.  For
simplicity, we have re-binned the XRT photon data to match the ROTSE-III
optical integration times.  For each of the optical integrations with
overlapping XRT data we have calculated the X-ray count rate.  We then
converted this count rate to $\fnux$ at 2.77 keV using the average afterglow
spectral parameters used to make the X-ray light curves as described in
\S~\ref{sec:xrt}.  We have neglected variations in the X-ray spectral index.
These are likely to be quite small for the following two reasons.  First, there
is no evidence for significant spectral evolution in the X-ray light curve,
excluding the initial steep decline and X-ray flares.  Second, we have
calculated $\fnux$ at 2.77 keV, which is the weighted mean of the X-ray
emission in the 0.3-10 keV range assuming a spectral index $\betax = -1.0$.
This ensures that slight changes in the X-ray spectral index from this
canonical value will not significantly alter $\fnux$.  We also note that we
have neglected any $k$-corrections, as these are impossible to calculate for
our unfiltered optical data.  However, we expect that the spectral index is
similar in the optical and X-ray bands.  Thus, the sense of the $k$-correction
will be the same for both X-ray and optical data, and may be neglected for
the purposes of this analysis.

\begin{figure*}
\begin{center}
\scalebox{0.7}{\rotatebox{270}{\plotone{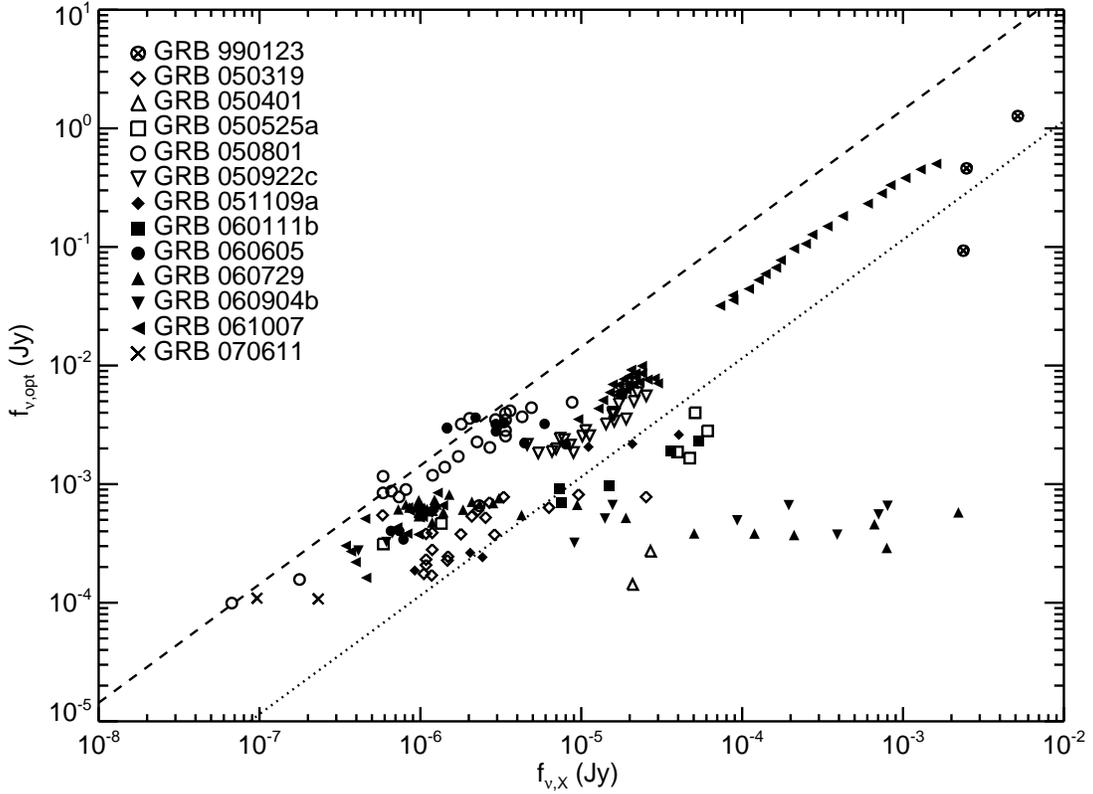}}}
\caption{\label{fig:fnuopt_fnux}Optical flux density $\fnuo$ [1.93 eV]
  vs. X-ray flux density $\fnux$ [2.77 keV] for 12 ROTSE-III bursts, as well as
  GRB~990123.  Each individual point in the Figure is from a specific burst at
  a single optical integration.  The size of the error bars are typically
  $\lesssim 10\%$ for the optical data, and $\lesssim 30\%$ for the X-ray data;
  the bright afterglow of GRB~061007 (leftward triangles) has very large
  signal-to-noise, and the error bars are smaller than the data points.  As an
  afterglow fades, the points will follow a track from the upper-right to the
  lower-left.  The dashed line shows the prediction from the synchrotron model
  where the optical and X-ray emission are in the same synchrotron regime with
  $\betaox = -1.0$.  The dotted line shows the same model with the cooling
  frequency at $0.3\,\mathrm{keV}$, just below the X-ray band.  With the
  exception of GRB~050401 (open upward triangles), the other observations below
  the dotted line all correspond to X-ray flares or the tail of the prompt
  $\gamma$-ray emission.}
\end{center}
\end{figure*}

Figure~\ref{fig:fnuopt_fnux} shows the optical flux density $\fnuo$ (at 1.93
eV) vs. the X-ray flux density $\fnux$ (at 2.77 keV).  Each individual point in
the figure is from a specific burst at a single optical integration.  The size
of the error bars are $\lesssim 10\%$ for the optical data and $\lesssim 30\%$
for the X-ray data.  As an afterglow fades, the points will follow a track from
the upper-right (bright in X-rays and optical) to the lower-left (faint in
X-rays and optical).  The dashed line shows the prediction from the synchrotron
model where the optical and X-ray emission are in the same synchrotron regime
($\nu_c < \nu_{\mathrm{opt}} < \nu_{\mathrm{X}}$) with $p=2.0$ and $\betaox =
-1.0$.  The dotted line shows the same model with $\nu_c = 0.3\,\mathrm{keV}$,
just below the X-ray band.  Most of the afterglow detections are between the
dashed and dotted lines.  As $\fnuo$ and $\fnux$ track each other, the spectra
are generally consistent with the predictions of the synchrotron model.
Furthermore, although there are many optical detections that are
``underluminous'' compared to the X-ray detections (those in the lower-right
corner of the plot), there are no optical detections that are significantly
``overluminous'', above the dashed line.  Each of these points is addressed in
turn.

There are two primary reasons for the optical emission to be underluminous in
Figure~\ref{fig:fnuopt_fnux}.  First, there might be significant extinction in
the host galaxy.  For example, this appears to be the case for
GRB~050401~\citep{dbbbb06}; we discuss the effects of local extinction in more
detail below.  Second, in the case of X-ray flares or the steep initial
decline, the X-ray flux may be dominated by internal shock emission, and thus
is not directly related to the optical flux.  This is the most likely
explanation for most of the underluminous optical detections in the lower-right
corner of the figure.  GRB~060729 (solid upward triangles) and GRB~060904b
(solid downward triangles) each have very bright X-ray flares, and GRB~050525a
(empty squares) has a shallow flare.  Meanwhile, flaring is not observed in the
mostly flat contemporaneous optical afterglows.  Excluding the duration where
there are obvious X-ray flares, these afterglows have observations that are
consistent with the main locus of points that falls within the range expected
from a simple synchrotron spectrum.  Similarly, other afterglows that appear to
be optically underluminous at the earliest times have early X-ray emission
dominated by a steep decline (e.g. GRB~051109a, solid diamonds) or prompt
emission (e.g. GRB~060111b, solid squares).  These are also consistent with a
simple broadband spectrum at later times.  This is another line of evidence
that the steep X-ray decay and X-ray flares are caused by internal shock
emission, and are not directly related to the external shock as traced by the
optical afterglow.

If the early optical light curve were caused by reverse shock
emission~\citep{mr97,sp99,sp99b,k00}, the optical emission may be
\emph{overluminous}.  A reverse shock is predicted to cause a prompt optical
flash, which will significantly outshine the optical forward shock emission
until the reverse shock crosses the ejecta shell.  Thus, if the contemporaneous
X-ray emission traces the forward shock, the optical emission from the reverse
shock will be brighter than an extrapolation of the forward shock synchrotron
spectrum.  A reverse shock has been hypothesized as the origin of early optical
emission for only a few afterglows, notably GRB~990123~\citep{abbbb99}.  The
primary evidence is the temporal evolution of the optical afterglow, which was
consistent with predictions: a fast rise followed by a steep ($\alphao \sim
-2$) decay and a break to a shallower decay ($\alphao \sim -1$).  However, a
similar temporal profile has not been observed for the vast majority of bursts
detected since GRB~990123, including the 12 bursts described in this paper.  In
addition, as shown in Figure~\ref{fig:fnuopt_fnux}, none of the bursts in this
paper require a separate optical component in excess of the predictions of the
external forward shock.  The implication that reverse shock emission is not
common has been noted by other authors~\citep[e.g.][]{mmkgg08,gkmgm09}.

We now investigate how the broadband spectral characteristics of the prompt
optical flash of GRB~990123, detected by ROTSE-I, compare to the 12 afterglows
in this paper.  The Wide Field Camera (WFC) on the \emph{Beppo}SAX satellite
obtained 2-10 keV X-ray observations of the prompt and early X-ray afterglow of
GRB~990123, contemporaneous with the prompt optical
flash~\citep{mmpfg05,cpkaa05}.  \citet{cpkaa05} performed spectral fits to the
WFC observations during the first three ROTSE-I integration times, including
the $9^{th}$ mag optical peak.  The X-ray flux densities $\fnux$ (2.77 keV)
obtained from the WFC observations are directly comparable to the XRT
observations taken at similar times for the ROTSE-III bursts.  In
Figure~\ref{fig:fnuopt_fnux} we have plotted the ROTSE-I $\fnuo$ vs. the WFC
$\fnux$ with the plotting symbol $\otimes$.  Although the optical flash from
GRB~990123 was exceptionally bright in optical, it was also exceptionally
bright in X-rays, comparable to GRB~061007.  However, unlike the early X-ray
afterglows detected for ROTSE-III bursts with a typical spectral index of
$\betax\sim-1.0$, GRB~990123 had a very hard X-ray spectrum, with a spectral
index ranging from $\betax\sim0$ to $\betax\sim0.5$.  Therefore, an
extrapolation of the X-ray spectrum to the optical regime greatly
\emph{underpredicts} the optical flux~\citep[see Fig. 2 in][]{cpkaa05}.
Although the broadband spectral index ($\betaox$) of the prompt optical flash
from GRB~990123 does not look quantitatively different from the ROTSE-III
detected afterglows, in this case the optical emission is significantly in
excess of an extrapolation of the X-ray emission.  This extra optical emission
component may be reverse shock emission, although other models have been
sugested, such as large-angle burst emission~\citep{pk07}.

We note that if the early X-ray observations are dominated by prompt emission,
and not the forward shock, then this simple broadband spectral analysis might
not hold.  This is most likely the case for GRB~060111b, where the rapidly
fading optical counterpart was detected \emph{prior} to the second peak of the
$\gamma$-ray emission, and for GRB~990123, where the spectrum of the early
X-ray emission was consistent with the GRB model function measured in the
$\gamma$-rays.  Nevertheless, it is remarkable that none of the optical
afterglow detections in this paper, at any time, are significantly brighter
than an extrapolation of the X-ray flux using a broadband power-law index
$\betaox = -1.0$.

The afterglows which have multiple observations that intersect the dashed line
in Figure~\ref{fig:fnuopt_fnux} merit particular attention.  These are
GRB~050801 (empty circles), GRB~060605 (solid circles), and GRB~070611
($\times$'s).  We neglect GRB~061007 (leftward triangles), for which a single
late time optical detection at is an anomalous outlier (see
Fig.~\ref{fig:combo3}). The peculiar GRB~050801 was been discussed in detail in
\citet{rmysa06}.  For this afterglow, the optical emission tracks the X-ray
emission over more than two orders of magnitude in time.  In addition, direct
extrapolation of the X-ray spectrum predicts $\fnuo$, indicating that the
optical and X-ray bands are in the same spectral regime.  The flux of the two
other optical afterglows along the dashed line, GRB~060605 (in the decay
phase), and GRB~070611 (in the final ROTSE-III observation), can also be
predicted by a direct extrapolation of the X-ray spectrum.  We note that the
X-ray spectra of these bursts do not show evidence for equivalent neutral
hydrogen absorption ($n_H$) in excess of Galactic.  Assuming the excess $n_H$
is attributable to the local environment, these afterglows will have minimal
local extinction.  Thus, for a given $\fnux$, these afterglows represent the
brightest possible $\fnuo$ after correcting for Galactic extinction and IGM
absorption.  No other optical afterglows are brighter than the simple
extrapolation of the X-ray spectrum.  This lends further support to our
hypothesis that the optical afterglow is dominated by the forward shock, even
at the earliest times.

\subsection{The Effect of Local Extinction}
\label{sec:extinct}

We now briefly address how local extinction in the host galaxy may cause a
given optical afterglow to appear underluminous.  GRB~050401 had a remarkably
dim afterglow, as well as a very large excess $n_H$ in the X-ray spectrum, as
shown in Table~\ref{tab:betax}.  Although local extinction has been posited as
the explanation for the dim afterglow, using the typical Milky Way (MW)
dust-to-gas ratio implies 20-30 mag of extinction, which results in an
unphysically bright optical counterpart~\citep{dbbbb06}.  Using well-measured
multi-band observations of optical afterglows, \citet{clw06} determined that no
single extinction law can plausibly explain the optical afterglow light curves,
and some bursts are well explained by a ``gray'' (flat) extinction law, rather
than a typical local extinction law.  In an expanded analysis, \citet{llw08}
used optical and X-ray afterglows that are in the same synchrotron regime to
constrain the extinction law as well as the dust-to-gas ratio for each burst.
They do not find a common connection between the optical extinction and X-ray
absorption.  Similarly, \citet{smpdm08} use UVOT and XRT data and find a wide
range in implied dust-to-gas ratio among the afterglows.

With our present single-filter ROTSE-III data, it is not possible to perform a
similar analysis to that performed by \citet{llw08} or \citet{smpdm08}.
However, we note that a cursory analysis hints that there is a loose relation
between $n_H$ as determined in the X-ray spectrum and the optical extinction.
Three of the afterglows with no excess column density (GRB~050801, GRB~060605,
and GRB~070611) have evidence for an absence of local optical extinction.  The
two afterglows with the largest excess column density (GRB~050401 and
GRB~051109a) are consistent with the largest optical extinction.  We leave it
to future work with multi-band optical data to explore this relationship further.

\section{Temporal Evolution}
\label{sec:temporal}

In \S~\ref{sec:spectra} we showed that the early optical afterglow and the
early X-ray afterglow generally follow the same track, after excluding the
optical rise and X-ray flares.  In order to remain on the same track, the
optical and X-ray afterglows must be fading at roughly the same rate.  In this
section we examine the temporal evolution of the afterglows in more detail.

\subsection{Power-law Fits}
\label{sec:plawfits}

In order to ascertain the gross temporal profile of the optical and X-ray
afterglows in the sample, we perform simple power-law fits to the light
curves.  These fits are intended to trace the overall time structure and not
the short timescale variability that is observed in some of the afterglows.

We fit a broken power-law to each of the optical and X-ray afterglows with the
following form:
\begin{eqnarray}
\label{eqn:bknpow}
f &  =  & a \left [ \left ( \frac{t}{t_b} \right )^{-s\alpha_1} + \left ( \frac{t}{t_b}
  \right )^{-s\alpha_2} \right ]^{1/s}\\
& = & \left ( \frac{t}{t_b} \right
  )^{\alpha_1} \left [ 1 + \left ( \frac{t}{t_b} \right
  )^{-s(\alpha_2-\alpha_1)} \right ]^{-1/s},
\end{eqnarray}
where $f$ is the flux, $t_b$ is the break time, $\alpha_1$ and $\alpha_2$ are
the two power-laws, and $s$ is a smoothing parameter.  We fix $|s|=10$ to yield
a moderately smooth break, where the sign of $s$ is positive when the light
curve is steepening and negative with the light curve is getting more shallow.
For the majority of light curve breaks, we do not have sufficient temporal
coverage or sensitivity to allow us to fit $s$.

Table~\ref{tab:optplaws} shows the results of the power-law fits to the optical
light curves.  For a few bursts there is ambiguity as to whether a single
power-law fit or a broken power-law fit is more appropriate; for completeness,
we have presented both fits in the table.  For the afterglow of GRB~060605, the
fit requires a double-broken power-law, which is a natural extension of
Eqn.~\ref{eqn:bknpow}.  Table~\ref{tab:xrtplaws} shows the results of the
power-law fits to the X-ray light curves, excluding X-ray flares, as noted in
the table.  We have also marked when there is ambiguity about the
preferred power-law model.  We note that some of the $\chi^2$ values for the
optical fits, in particular, are quite poor.  For example, the optical
afterglow of GRB~061007 shows short timescale variability which leads to a very
large $\chi^2$ for the fit, although a single power-law with decay index
$-1.66\pm0.01$ is quite effective at describing the overall trend in the
optical decay for the duration of the ROTSE-III observations.

\begin{deluxetable*}{ccccccc}
\tablewidth{0pt}
\tablecaption{Power-law fits to ROTSE-III data\label{tab:optplaws}}
\tablehead{
\colhead{GRB} &
\colhead{Fit $t_\mathrm{start}$ (s)} &
\colhead{Fit $t_\mathrm{stop}$ (s)} &
\colhead{$\alpha$} &
\colhead{$t_\mathrm{break}$ (s)} &
\colhead{$\chi^2/\nu$} &
\colhead{In Fig~\ref{fig:alphacompare}}
}
\startdata
GRB~050319 & 169 & 5000 & $-0.89\pm0.03$ & n/a & 52.4/32 & *\\
GRB~050401 & 35 & 241 & $-0.69\pm0.18$ & n/a & 3.2/3 & *\\
GRB~050525a & 406 & 10843 & $-0.31\pm0.07$ & $4100\pm350$ & 38.8/19 & *\\
           &     &       & $-1.27\pm0.16$ & -- & & *\\
GRB~050801 & 22 & 10000 & $-0.12\pm0.01$ & $228\pm6$ & 116/42 & *\\
          &    &       & $-1.10\pm0.01$ & -- & & *\\
GRB~050922c\tablenotemark{a} & 174 & 3630 & $-0.74\pm0.02$ & n/a & 46.4/23\\
GRB~050922c\tablenotemark{b} & 174 & 3630 & $-1.18\pm0.14$ & $364^{+109}_{-58}$ & 8.9/21 & *\\
           &     &      & $-0.66\pm0.03$ & -- & & *\\
GRB~051109a & 39 & 13300 & $-0.65\pm0.01$ & n/a & 278.5/38 & *\\
GRB~060111b & 35.3 & 179 & $-2.35\pm0.10$ & n/a & 8.0/7 & *\\
GRB~060605 & 74 & 6317 & $1.18\pm0.33$ & $152\pm25$ & 55.5/48 & *\\
          &    &      & $0.14\pm0.06$ & $666\pm32$ & & *\\
          &    &      & $-1.00\pm0.03$ & -- & & *\\
GRB~060729 & 306 & 3016 & $0.91^{+0.67}_{-0.49}$ & $424^{+79}_{-45}$ & 50.4/39 &
*\\
          &     &      & $-0.20\pm0.04$ & -- & & *\\
GRB~060904b\tablenotemark{a} & 1694 & 6440 & $-0.44\pm0.06$ & n/a & 11.5/7 & *\\
GRB~060904b\tablenotemark{b} & 583 & 6440 & $-1.8\pm0.4$ & $870\pm80$ & 38.2/21\\
           &     &      & $-0.25\pm0.04$ & -- & \\
GRB~061007 & 108 & 14600 & $-1.66\pm0.01$ & n/a & 390/78 & *\\
GRB~070611 & 768 & 8900 & $2.1\pm0.6$ & $2230\pm200$ & 7.1/8\\
          &     &      & $-0.61\pm0.14$ & -- & & *\\
\hline
\enddata
\tablenotetext{a}{Single power-law fit.}
\tablenotetext{b}{Broken power-law fit.}
\end{deluxetable*}

\begin{deluxetable*}{ccccccc}
\tablewidth{0pt}
\tablecaption{Power-law fits to XRT data\label{tab:xrtplaws}}
\tablehead{
\colhead{GRB} & 
\colhead{Fit $t_\mathrm{start}$ (s)} &
\colhead{Fit $t_\mathrm{stop}$ (s)} &
\colhead{$\alpha$} &
\colhead{$t_\mathrm{break}$ (s)} &
\colhead{$\chi^2/\nu$} &
\colhead{In Fig~\ref{fig:alphacompare}}
}
\startdata
GRB~050319 & 240 & 13400 & $-4.7^{+0.7}_{-1.1}$ & $410\pm30$ & 6.0/11 &\\
          &     &       & $-0.51\pm0.05$ & -- & & *\\
GRB~050401 & 143 & 19843 & $-0.59\pm0.02$ & $4670\pm700$ & 197.2/187 & *\\
          &     &       & $-1.37\pm0.11$ & -- & \\
GRB~050525a\tablenotemark{a} & 130 & 58537\tablenotemark{c} & $-1.25\pm0.01$ & n/a & 22.2/22 & *\\
GRB~050525a\tablenotemark{b} & 130 &  58537 & $-0.62\pm0.02$ & $1040\pm80$ & 95.1/90 & *\\
           &   &        & $-1.71\pm0.04$ & -- & & *\\
GRB~050801 & 70 & 50000 & $0.04^{+0.5}_{-0.3}$ & $270^{+70}_{-50}$ & 9.7/17 & *\\
          &    &       & $-1.16\pm0.05$ & -- & & *\\
GRB~050922c\tablenotemark{a} & 117 & 65816 & $-1.17\pm0.01$ & n/a & 108.4/75 & \\
GRB~050922c\tablenotemark{b} & 117 & 65816 & $-0.74\pm0.16$ & $289^{+132}_{-61}$ & 74.1/73 & *\\
           &     &       & $-1.23\pm0.03$ & -- & & *\\
GRB~051109a\tablenotemark{a} & 129 & 200 & $-3.1\pm0.4$ & n/a & 4.2/11 & \\
GRB~051109a\tablenotemark{a} & 3500 & 25000 & $-1.03\pm0.05$ & n/a & 21.5/36 & *\\
GRB~060111b & 93 & 53058 & $-4.64^{+0.8}_{-1.1}$ & $129\pm10$ & 20.4/22 & *\\
           &    &       & $-1.09\pm0.03$ & -- & &\\
GRB~060605 & 102 & 60000 & $-1.6\pm0.6$ & $231^{+153}_{-21}$ & 15.4/18 & *\\
          &     &       & $-0.34\pm0.08$ & $5770\pm600$ & & *\\
          &     &       & $-1.89\pm0.08$ & -- & & *\\
GRB~060729 & 200 & 12181 & $-7.6\pm0.14$ & $401\pm11$ & 89/91 & \\
          &     &       & $-1.86\pm0.6$ & $687\pm80$ & & *\\
          &     &       & $0.0\pm0.03$ & -- & & *\\
GRB~060904b\tablenotemark{a} & 3600 & 37000 & $-1.37\pm0.06$ & n/a & 10.8/14 & *\\
GRB~060904b\tablenotemark{b} & 89 & 37000\tablenotemark{d} & $-0.76\pm0.04$ & $5600\pm1500$ & 9.5/14\\
           &    &       & $-1.49\pm0.12$ & -- &\\
GRB~061007 & 87 & 24400 & $-1.68\pm0.01$ & n/a & 540/322 & *\\
GRB~070611 & 3392 & 42000 & $-0.84\pm0.23$ & n/a & 5.2/8 & *\\
\enddata
\tablenotetext{a}{Single power-law fit.}
\tablenotetext{b}{Broken power-law fit.}
\tablenotetext{c}{Excluding the flare from 150 s--300 s}
\tablenotetext{d}{Excluding the small flare from 200 s--1000 s}
\end{deluxetable*}

The X-ray light curves show the steep-shallow-steep morphology with
superimposed flares that has been referred to as the ``canonical'' X-ray
afterglow~\citep{nkgpg06}.  The optical afterglows do not have such an obvious
pattern.  For example, we do not always see the rise of the optical afterglow,
even when the first ROTSE-III image is taken within seconds after the start of
$\gamma$-ray emission.  The implications of the variety in optical rise time
are addressed in \S~\ref{sec:risetimes}.  After the initial rise, some of the
optical afterglows decay very slowly (e.g. GRB~050801, $\alpha\sim -0.1$) and
some very rapidly (e.g. GRB~061007, $\alpha\sim -1.7$).  For most of the
optical afterglows where we see an early break, the initial decay is shallower
than the later decay, although there are exceptions, such as GRB~050922c (and
GRB~990123) which show evidence for an initial shallowing rather than
steepening of the light curve.

With the power-law fits, we can also begin to constrain the break times in the
optical and X-ray light curves.  The typical explanations for a light curve
break at the early time are as follows.  First, if the fireball is expanding
into a constant density medium, when the synchrotron cooling frequency,
$\nu_c$, passes through the optical waveband, the decay index should steepen by
$\delta\alpha=0.25$, without a contemporaneous change in the X-ray decay index.
Second, cessation of long duration energy injection should cause an identical
steepening of both the optical and X-ray decays, with the magnitude arbitrarily
determined by the rate of energy injection.  Third, a sudden change in the
density of the surrounding medium may cause a chromatic or achromatic break,
depending on whether the observed band is above or below $\nu_c$.  Finally, if
the early optical afterglow is dominated by reverse shock emission, we expect a
steep-to-shallow transition in the decay index.

For several \swift{} afterglows, previous work has shown that many light curve
breaks are chromatic, and are not observed simultaneously in both the optical
and X-ray wavelengths~\citep[e.g.][]{pmbng06,mmkgg08}.  These chromatic breaks
are difficult to reconcile with the hypothesis that the flat portion of the
X-ray afterglow is caused by continuous energy injection: when the energy
injection stops, we should see a break in all bands simultaneously.  The
magnitude of the breaks are also typically not consistent with the passage of
$\nu_c$, or with a change in density of the circumburst medium.  Using
additional optical data, the study by \citet{pmbng06} documents unexplained
chromatic breaks in three of the afterglows reported in this paper: GRB~050319,
GRB~050401, and GRB~059022c.  On the other hand, some afterglows do show
achromatic breaks at the early time~\citep[e.g. GRB~050801, as described
in][]{rmysa06}.  Unfortunately, the ROTSE-III telescopes are only able to
detect afterglows for a limited duration, and there are orbital gaps inherent
in the XRT coverage.  Combined, this makes it difficult in this work to shed
further light on this important topic. We leave it to future work to integrate
the ROTSE-III light curves with later optical observations (as had been done
with GRB~050319 and GRB~050401) to better constrain the precise timing of the
optical and X-ray breaks.

\subsection{Optical and X-ray power-law comparison}
\label{sec:plawcompare}

\begin{figure}
\begin{center}
\scalebox{0.85}{\rotatebox{270}{\plotone{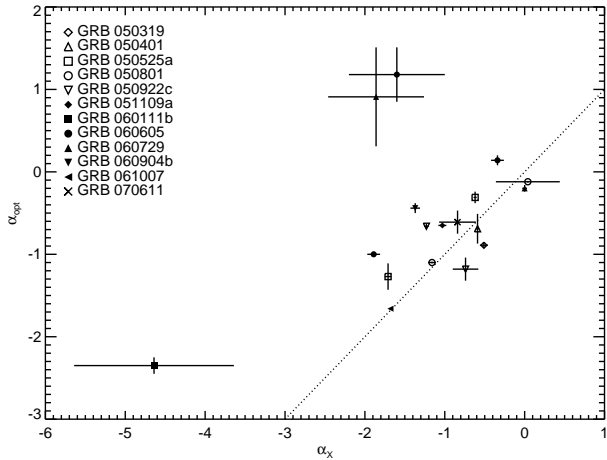}}}
\caption{\label{fig:alphacompare}Power-law indices $\alphao$
  vs. $\alphax$ for the 12 ROTSE-III afterglows.  The dotted line is the line
  of equality.  In general, the afterglows that fade rapidly in optical fade
  rapidly in X-rays.  The notable exception are the X-ray flares (not shown)
  and the early optical rise (seen here in early observations of GRB~060605 and
  GRB~060729). The optical afterglows tend to fade slightly slower than the
  X-ray afterglows, as predicted if the forward shock is expanding into a
  constant density medium.}
\end{center}
\end{figure}

We now investigate the relationship between the optical and X-ray decay rates
($\alphao$ and $\alphax$) for each afterglow in the sample.  These two
quantities are plotted in Figure~\ref{fig:alphacompare}.  Each data point
represents a single contemporaneous determination of the temporal index
$\alpha$ for the ROTSE-III and XRT observations, as denoted by asterisks in
Tables~\ref{tab:optplaws} and \ref{tab:xrtplaws}.  If there were no overlapping
observations of a particular temporal segment (e.g. for the optical rise of
GRB~061007) then there is no data point on the plot.  The dashed line of
equality is shown for reference.

Although there is a large scatter, we observe a significant trend in the data
indicating that the more rapidly fading X-ray afterglows are correlated with
rapidly fading optical afterglows.  We note that this tracking behavior was
hinted at in the broadband spectral comparison in Figure~\ref{fig:fnuopt_fnux}.
The three largest outliers are GRB~060111b, where the early optical afterglow
is contemporaneous with the tail of the prompt emission observed in X-rays; and
the early optical detections of GRB~060605 and GRB~060729, where the optical
rise ($\alphao > 0$) is paired with a steep decline in the X-rays. 

The trend in Figure~\ref{fig:alphacompare} is most evident in the extreme
cases.  For example, GRB~050801 (empty circle) and GRB~060729 (solid upward
triangle) both have flat decays that are contemporaneous in optical and X-rays.
At the lower left of the distribution, GRB~050801 (empty circle), GRB~061007
(leftward triangle), and GRB~050525a (empty square) both show quite steep
declines contemporaneously.  The rest of the decay indices are all clustered in
the middle, with $\alphax \sim -1$ and $\alphao \sim -0.5$.


The standard fireball model for the afterglow predicts that when the forward
shock expands into a constant density (ISM) medium, $\alphax \propto
t^{(2-3p)/4}$ and $\alphao \propto t^{3(1-p)/4}$ when the synchrotron cooling
break $\nu_c$ is between the optical and X-ray regime~\citep{gs02}.  Taking our
fiducial electron index of $p=2$ that is implied from the typical X-ray
spectral index ($\betax \sim -1.0$, with $\betax = -p/2$), then we predict
$\alphax = -1.0$ and $\alphao = -0.75$.  On the other hand, if the shock is
expanding into a $n \propto 1/r^2$ wind medium, then we predict $\alphax =
-1.0$ and $\alphao = -1.25$.  When $\nu_c$ is in between the X-ray and optical
regime, we expect the optical decay will be slightly slower than the X-ray
decay for the ISM case, and we expect the inverse in the wind case. If the
X-ray and optical observations sample the same spectral regime (ie., the the
cooling break is below the optical regime), we expect $\alphax = \alphao \sim
-1.0$ for both the ISM and wind cases.

It is challenging to shoehorn individual afterglows into this simple toy model.
However, in the aggregate they are roughly consistent with the predictions for
the forward shock expanding into an ISM medium.  First, we observe that the
optical and X-ray afterglows typically track each other.  Second, the optical
decay index $\alphao$ is usually \emph{slightly} shallower than the X-ray
decay index $\alphax$.  Meanwhile, two well-sampled bursts, GRB~050801 and
GRB~061007, have the same optical and X-ray decay indices, as also indicated in
Figure~\ref{fig:fnuopt_fnux}.  In these cases, the optical and X-ray
observations are consistent with sampling the same spectral regime.  Two
afterglows, GRB~050319 (empty diamonds) and GRB~060729 (solid upward triangles)
decay more rapidly in the optical than the X-rays.  This may indicate a
different external medium, or it may just represent the large scatter in
the $\alphax-\alphao$ correlation.

Although the difference between the optical and X-ray decay indices is roughly
consistent with the predictions of the fireball model, the magnitude of the
decay indices are not.  Thus, using the X-ray spectral index $\betax$ to infer
the electron index $p$ yields different results than using the afterglow decay
index $\alphax$ or $\alphao$ to infer $p$.  These discrepancies are most
apparent in the early afterglows of GRB~060729 and GRB~061007.  The
extraordinarily flat and long-lived afterglow of GRB~060729 has been
interpreted as a sign of long-duration energy injection~\citep{ggwrc07}, and
the X-ray detection almost two years after the burst may also imply a very wide
jet~\citep{gbxwz09}.  On the other side, the remarkably steep decay of
GRB~061007 with no indication of a late jet break (or any other break) has been
interpreted as evidence for a strongly collimated jet,~\citep{sdpvp07}.  We
note that in our present analysis there is nothing significantly different
between these afterglows at the early times except for the absolute value of
the decay indices.  Thus, in order to obtain a coherent picture of the early
afterglow emission, we must (a) posit a different emission model for each of
these bursts individually, or (b) modify the afterglow modeling such that we
can accommodate a standard spectral index and a wide range of temporal
indices.

We emphasize again that the early optical and X-ray afterglows, after excluding
the steep decay phase and X-ray flares, appear to be probing the same emission
region.  Furthermore, the behavior is roughly consistent with the model for the
forward shock.  We have assumed that the X-ray spectral index, which is common
among the set of afterglows, is a more robust tracer of the underlying electron
index.  The absolute value of optical and X-ray decay indices shows much more
variety, and depends much more strongly on the nature of the circumburst
environment and posited long-duration energy injection. 



\section{Optical Rise Times}
\label{sec:risetimes}


Due to their rapid response capabilities and sensitivity to typical early
afterglows, the ROTSE-III telescopes are uniquely suited to measure the optical
rise time of the typical GRB optical counterpart.  The optical rise may be
tracing the onset of the forward shock~\citep[e.g. GRB~060418, ][]{mvmcd07} or
the reverse shock~\citep[e.g. GRB990123, ][]{abbbb99}, and may be combined with
optical emission correlated with the prompt $\gamma$-rays from internal
shocks~\citep[e.g. GRB~050820a, ][]{vwwag06}.  As we have shown, the broadband
early afterglow spectra of the bursts described in this paper are all
consistent with forward shock emission.  In this case, the peak of the optical
afterglow can constrain the initial bulk Lorentz factor of the GRB
ejecta~\citep{sp99}.

Using the REM telescope, \citet{mvmcd07} detected the optical rise of two early
afterglows, GRB~060418 and GRB~060607a, and were able to constrain the initial
bulk Lorentz factor of the outflow.  Both of these early optical afterglows
followed a very smooth evolution in the early rise and fall, with a peak
$\sim150\,\mathrm{s}$ after the burst.  Furthermore, REM obtained imaging of
GRB~060418 with multiple filters, and did not detect significant color
evolution at the early time.  This is consistent with the predictions for the
onset of the forward shock~\citep{sp99}.

The peak time of the optical afterglows of both GRB~060418 and GRB~060607a
occurred after the end of the $\gamma$-ray emission, such that $\tpk > T_{90}$.
Therefore, the outflow can be modeled as a ``thin shell,'' the definition of
which depends on the relationship between the thickness of the outflowing shell
in relation to its Lorentz factor~\citep{sp99}.  The ``thick shell'' case,
where the forward shock peak occurs during the $\gamma$-ray emission, is more
difficult to model.  In the thin-shell case, the peak of the optical emission
$t_\mathrm{pk}$ corresponds to the deceleration timescale $t_\mathrm{dec} \sim
R_\mathrm{dec}/(2c\Gamma^2_\mathrm{dec})$, where $c$ is the speed of light,
$R_\mathrm{dec}$ is the deceleration radius, and $\Gamma_\mathrm{dec}$ is the
bulk Lorentz factor of the fireball at $t_\mathrm{dec}$.  The initial Lorentz
factor $\Gamma_0$ is expected to be twice the Lorentz factor at the
deceleration time, $\Gamma_\mathrm{dec}$~\citep{m06}.  Assuming that the
fireball is expanding in a homogeneous medium with constant particle density
$n$, consistent with the conclusions of \S~\ref{sec:plawcompare}, we find
\begin{eqnarray}
\Gamma_0 \sim 2\Gamma_\mathrm{dec} &  = & 2 \left [ \frac{3\eiso}{32\pi n
    m_\mathrm{p} c^5 \eta t_{\mathrm{pk},z}^3} \right ]^{1/8}\\
 & \approx & 560 \left
      [ \frac{\eisoft}{\eta_{0.2} n_0 t_{\mathrm{pk},z,10}^3} \right ]^{1/8},
\end{eqnarray}
where $\eisoft$ is the isotropic equivalent energy release in
$\gamma$-rays in units of $10^{52}\,\mathrm{ergs}\,\mathrm{s}^{-1}$; $n$ is
local density and $n = n_0\,\mathrm{cm}^{-3}$; $m_p$ is the mass of the proton;
$\eta = 0.2 \eta_{0.2}$ is the radiative efficiency; and $t_{\mathrm{pk},z,10}
= \tpk/[(1+z)\times 10\mathrm{s}]$ is the optical peak-time corrected
for cosmological time dilation in units of $10\,\mathrm{s}$.  

The calculation of $\Gamma_0$ requires an estimation of $\eiso$ for each of the
GRBs.  Unfortunately, BAT only covers a narrow band-pass of 15-150~keV.  Some
of the bursts listed in this paper have been observed by Konus-Wind, which has
a much wider band-pass more suitable for measuring the peak of the $\gamma$-ray
emission, $\epk$, and thus allowing an estimation of $\eiso$.  As high energy
coverage is limited, \citet{bkbc07} have used a Bayesian analysis using priors
from BATSE spectra to estimate $\epk$ and $\eiso$ directly from BAT data.  As
this analysis can be consistently applied to all of the bursts in this paper,
we have chosen to use $\eiso$ from \citet{bkbc07}.  For the GRBs with
Konus-Wind data, we have confirmed that the estimates of $\eiso$ are consistent
for each of the methods.  Furthermore, the estimation of $\Gamma_0$ is only
weakly dependent on $\eiso$ ($\Gamma_0 \propto \eiso^{1/8}$), such that an
error of factor of 2 in $\eiso$ will only shift $\Gamma_0$ by $\sim 10\%$.

\subsection{Calculating the Optical Peak Time}
\label{sec:calcpk}

We have calculated the time of the optical peak, $\tpk$, from the early
ROTSE-III data for each of the bursts in this paper.  For seven of the bursts,
the afterglow was already fading at the initial ROTSE-III observation, and thus
we only have an upper limit on the time of the optical peak.  For the
afterglows for which we observe the optical rise, we have fit a smoothly broken
power-law of the form:
$$
f=\left ( \frac{t}{t_b} \right )^{\alpha_1} \left [ 1 + \left ( \frac{t}{t_b}
  \right )^{-s(\alpha_2 - \alpha_1)} \right ]^{-1/s},
$$
where $f$ is the flux, $t_b$ is the break time, $\alpha_1$ and $\alpha_2$ are
the two power-law indices, and $s$ is a smoothing parameter.  When $s>0$ then
this function can fit the light curve peak.  We can then calculate the peak
time $t_p$:
$$
t_p = t_b \left ( \frac{-\alpha_1}{\alpha_2} \right
)^{1/[s(\alpha_1-\alpha_2)]}.
$$ When performing the fits, we allow all parameters to float, including the
smoothing parameter, with the constraint that $0.01<s<50$.  For the afterglows
where $s$ could not be constrained, our error bars are essentially marginalized
over all values of $s$.  Unlike in the case with the power-law fits in the
decaying phase of the afterglow (see \S~\ref{sec:plawfits}), we can fit for $s$
due to the increased leverage in the rising and fading afterglow.  Furthermore,
we note that we only perform the fit for a suitable time interval around the
peak of the burst, to limit the contamination from later light curve breaks.

The results of the fits are shown in Table~\ref{tab:optpkfits}.  For two of the
afterglows, GRB~060729 and GRB~060904b, there appear to be two optical peaks,
and we have listed both peak times in the table.  With GRB~060729 (see
Figure~\ref{fig:combo3}), there is an initial optical flare which may be
correlated with on of the peaks in the prompt $\gamma$-ray emission; the smooth
shape of the later light curve peak at $\tpk = 485\,\mathrm{s}$ is more
suggestive of the onset of the forward external shock.  With GRB~060904b (see
Figure~\ref{fig:combo3}), there appears to be two optical peaks, although
neither is correlated with the high energy emission.  The smooth profile of the
later peak is more suggestive of the onset of the forward shock.  For all the
bursts, in addition to the question of which optical peak is the onset of the
external shock, there may be an ambiguity of what is the proper time to use for
the start time, $t_0$.  This issue is discussed below.

\begin{deluxetable*}{cccccccc}
\tablewidth{0pt}
\tablecaption{Optical Peak Fits\label{tab:optpkfits}}
\tablehead{
\colhead{GRB} &
\colhead{$t_\mathrm{start} (s)$} &
\colhead{$t_\mathrm{stop} (s)$} &
\colhead{$\alpha_1$} &
\colhead{$\alpha_2$} &
\colhead{$s$} &
\colhead{$t_{pk} (s)$} &
\colhead{$\chi^2/dof$}
}
\startdata
GRB~060605 & 73 & 6317 & $0.40^{+0.13}_{-0.09}$ & $-1.02\pm0.05$ &
$3.4^{+2.2}_{-1.3}$ & $484\pm40$ & $1.3$\\
GRB~060729 &     &      &        &                        & &
$96$\tablenotemark{a} & \\  
          & 218 & 3016 & $>0.8$ & $-0.21^{+0.04}_{-0.09}$ & $4.3$\tablenotemark{b}
& $485^{+97}_{-39}$ & $1.2$ \\
GRB~060904b & 29.8 & 120 & $>0.2$ & $<-0.8$ & $2.7$\tablenotemark{b} &
$53^{+11}_{-8}$ & 0.04\\
 & 217 & 1485 & $1.7_{-0.7}^{+2.0}$ & $-1.11^{+0.14}_{-0.20}$ &
$2.4_{-1.7}^{+4.2}$ & $453^{+30}_{-15}$ & 1.5\\
GRB~061007 & 29.7 & 554 & $27.4\pm0.4$ & $-1.60\pm0.01$ & $<0.02$ &
$71.1\pm30$ & 47\tablenotemark{c}\\
GRB~070611 & 779 & 8867 & $2.0^{+0.7}_{-0.4}$ & $-0.61\pm0.14$ & $>5$ &
$2296\pm151$ & 0.976\\
\enddata
\tablenotetext{a}{There is insufficient coverage around the time of the short
  duration flare to perform a fit.}
\tablenotetext{b}{$s$ is unconstrained within the range $0.01<s<50$}
\tablenotetext{c}{This model is not a good fit to the early data; see text for
  details.}
\end{deluxetable*}

Special attention must be paid to the early afterglow of GRB~061007, as our
simple model is a poor fit.  There is significant structure in the early
afterglow, with two sub-peaks at $\sim60\,\mathrm{s}$ and
$\sim100\,\mathrm{s}$.  Furthermore, the initial optical decline is not quite a
power-law, with substructure that is visible due to the relatively high
precision photometry made possible by the brightness of the afterglow.
Although the model, which traces out the overall shape, suggests that the
afterglow has a peak time of $71\,\mathrm{s}$, we also consider the two
observations with maximum brightness to be valid peaks.  Therefore, we have
added a systematic error of $\pm30\,\mathrm{s}$ to the quoted peak time.

\subsection{Estimating $\Gamma_0$}
\label{sec:gamma0}

Under the assumption that the thin shell model approximations are valid, we can
now estimate the initial bulk Lorentz factors for the GRBs.  These are
tabulated in Table~\ref{tab:risetimes}.  For the bursts where the initial ROTSE
observation was obtained after the afterglow was already fading, we obtain a
lower limit in the bulk Lorentz factor.  For the remaining bursts, we see a
large range in initial Lorentz factors, ranging from relatively slow jets in
bursts such as GRB~070611 ($\Gamma_0 \sim 100$) to highly ultrarelativistic
jets in burst such as GRB~050401 ($\Gamma_0 \gtrsim 900$).  This latter burst
is notable in that it has the largest $\eiso$ in the set, combined
with one of the most rapid implied optical rise times, after correcting for
cosmological time dilation.  In combination, this implies a very fast jet.  We
also note that in no case is the initial bulk Lorentz factor estimated to be
smaller than $\Gamma_0 \sim 100$, which has been estimated as the minimum for
which the $\gamma$-ray production is possible~\citep{p05}.

At this stage, it is worth exploring the validity of our assumptions for
calculating the initial Lorentz factor, especially as it applies to individual
bursts.  These include: that the optical afterglow traces the forward external
shock; the validity of the thin shell approximation;  and that we have used the
proper start time, $t_0$.

We first address the assumption that the optical afterglow traces the forward
external shock.  As we have shown previously, the broadband early afterglow
spectra are all consistent with forward shock emission, with the notable
exception of the X-ray flares.  The optical afterglow appears to be less
contaminated by prompt (internal shock) emission than the X-ray afterglow, and
thus is a better tracer of the forward shock.  However, the unfiltered
ROTSE-III observations cannot constrain the color evolution in the early
afterglow; in other afterglows detected by multi-band instruments, this color
evolution appears to be modest or absent, consistent with the external shock
hypothesis~\citep{yaraa06,mvmcd07,pbbph08}.  It is not possible to make any
definitive statements about color evolution with our present ROTSE-III
data set, and therefore we cannot rule out the possibility that the early
afterglow does not have a significant reverse shock component.  In addition,
there are two afterglows, GRB~060729 and GRB~060904b, for which there appear to
be two optical peaks.  In each case there is a short-duration peak
contemporaneous with the $\gamma$-ray emission, and a smoother peak after the
end of the $\gamma$-ray emission.  Thus, it seems reasonable to assume the
latter optical peak traces the onset of the forward shock.

For the afterglows in which we detect the early rise, we can also check if the
power-law index of the rise is consistent with predictions.  As the forward
shock expands into a homogeneous ISM medium, the afterglow is expected to rise
as $\fnuo \propto t^2$~\citep{pv08}.  This is roughly consistent with the onset
index of GRB~070611 ($\alpha = 2.1\pm0.6$).  However, the rise of GRB~060605
($\alpha=1.18\pm0.33$) and GRB~060729 ($\alpha = 0.9^{+0.7}_{-0.5}$) are
shallower, and GRB~061007 ($\alpha \sim 9$) is significantly steeper.  As with
the afterglow decay, we find a large range of temporal indices, with a large
scatter around the predicted value.  Here, our conclusions are the same as
reported at the end of \S~\ref{sec:plawcompare}.  Since the broadband spectra
are consistent with the onset of the forward shock, we attribute the large
scatter in the temporal decay indices to the current simplicity of our model.

We next address the validity of the thin shell approximation.  As shown in
Table~\ref{tab:risetimes}, in all the cases except for GRB~060111b and
GRB~061007 the optical peak comes after the end of significant $\gamma$-ray
emission as measured by $T_{90}$.  Therefore, for the majority of bursts, and
especially those with very short durations ($T_{90}/z\sim2\,\mathrm{s}$) the
thin shell approximation should be valid.  GRB~061007 requires careful
attention.  For this burst, the optical peak occurs soon after the end of the
main $\gamma$-ray emission.  Thus, the modeling of this afterglow is certainly
more complicated, as demonstrated by the structure in the optical light curve
near its peak.  Due to the very smooth evolution of the later afterglow, which
is consistent with the forward shock, we can infer that the broad outline of
the optical peak traces the onset of the forward shock, perhaps with additional
contribution from prompt emission near the optical peak.  For GRB~060111b, we
do not have sufficient ROTSE-III coverage to fully constrain the light curve
evolution.  It is apparent that the optical peak occurs prior to the
second BAT peak (see Fig.~\ref{fig:combo2}), implying that the thin shell
approximation may not be valid in this case. 

\begin{deluxetable*}{cccccc}
\tablewidth{0pt}
\tablecaption{Optical Rise Times and Implied $\Gamma_0$\label{tab:risetimes}}
\tablehead{
\colhead{GRB} &
\colhead{z} &
\colhead{$t_\mathrm{pk}/(1+z)$} &
\colhead{$T_{90}/(1+z)$\tablenotemark{a}} &
\colhead{$\eiso$} &
\colhead{$\Gamma_O$}\\
& & $(s)$ & (s) &
$(10^{52}\,\mathrm{ergs}\,\mathrm{s}^{-1})$ &
$\times(\eta_{0.2}n_0)^{-1/8}$ 
}
\startdata
GRB~050319 & 3.24 & $<39$ & 36 & $4.6^{+6.5}_{-0.6}$ & $\gtrsim 400$\\
GRB~050401 & 2.9 & $<9$  & 8.5 & $32^{+26}_{-7}$ & $\gtrsim 900$\\
GRB~050525a & 0.61 & $<253$ & 5.7 & $2.04^{+0.11}_{-0.09}$ & $\gtrsim 180$\\
GRB~050801 & 1.6 & $<9$ & 7.5 & $0.22^{+0.36}_{-0.03}$ & $\gtrsim 500$\\
GRB~050922c & 2.2 & $<55$ & 1.4 & $3.9^{+2.7}_{-0.8}$ & $\gtrsim 350$\\
GRB~051109a & 2.35 & $<11$ & 11.1 & $2.3^{+2.4}_{-0.5}$ & $\gtrsim 600$\\
GRB~060111b & $\sim1.0$ & $\lesssim18$ & $\sim29$ & $2.9^{+5.2}_{-1.3}$ & $\gtrsim 500$\\
GRB~060605 & 3.8 & $101\pm8$ & 16.5 & $2.5^{+3.1}_{-0.6}$ & $\sim 260$\\
GRB~060729 & 0.54 & $\sim62$ & 75 & $0.33^{+0.29}_{-0.06}$ & $\sim 250$\\
          &      & $318^{+60}_{-27}$  & & $0.33^{+0.29}_{-0.06}$ & $\sim 133$\\
GRB~060904b & 0.70 & $31\pm6$ & 101 & $0.30^{+0.19}_{-0.06}$ & $\sim 310$\\
           &      & $282^{+24}_{-15}$ & & $0.30^{+0.19}_{-0.06}$ & $\sim 140$\\
GRB~061007 & 1.26 & $31\pm15$ & 33 & $140^{+110}_{-60}$ & $\sim680$\\
GRB~070611 & 2.04 & $755\pm50$ & 4 & $0.34\pm0.06$ & $\sim 100$\\
\enddata
\tablenotetext{a}{Values of $T_{90}$ taken from \citet{sbbcf08}.}
\end{deluxetable*}

Finally, we note that these calculations are sensitive to the calculation of
the start time of the burst, $t_0$.  As we have noted in \S~\ref{sec:bat}, the
quoted values of $t_0$, defined as the first detection of BAT emission with
SNR$>6$, have a typical error of $\pm5\,\mathrm{s}$.  This may be significant,
especially for the afterglows with the earliest limits, such as
GRB~050801. After correcting for cosmological time dilation, even this
relatively large uncertainty implies an uncertainty in $\Gamma_0$ of less than
10\%.  More problematic are $\gamma$-ray precursors that may not have been
detected by BAT due to coverage or sensitivity.  A particular cautionary tale
is GRB~050319.  The original analysis of early optical data showed a shallow
light curve break, when using the BAT trigger time~\citet{wvwwe05}.  Further
analysis of BAT slew data demonstrated that the main GRB event was in fact
137~s prior to the original BAT trigger time~\citet{cmrct05}.  As discussed in
\citet{qryaa06}, using a grossly incorrect $t_0$ can create spurious breaks in
the afterglow light curve, especially at small $t/\delta t$.  Similarly, an
incorrect calculation of $t_0$ will bias the estimation of the bulk Lorentz
factor $\Gamma_0$.  Although we do not know of any other bursts in this paper
that had additional episodes of $\gamma$-ray emission bright enough to be
detected by BAT, we of course cannot rule out the possibility that $t_0$ is not
correct.

\section{Discussion}
\label{sec:discussion}

The ROTSE-III telescope network has a unique combination of response time,
aperture, and global coverage to trace the evolution of optical afterglows from
the earliest time.  By combining ROTSE-III observations with the rapid response
capabilities of the \swift{} telescope, we can track the earliest phases of the
prompt optical emission and afterglow as soon as $10\,\mathrm{s}$ after the
start of the burst, and follow the afterglow for many orders of magnitude in
time.  In this paper we have assembled a complete set of 12 ROTSE-III afterglow
light curves observed between March 2005 and June 2007 with a significant
number of early optical observations and simultaneous coverage with XRT.  With
a median response time of $45\,\mathrm{s}$ from the start of $\gamma$-ray
emission (8~s after the GCN notice time), we have the opportunity for a unique
look at the onset of the afterglow emission.

This paper has attempted to focus on the commonalities among the set of 12
afterglows, rather than the differences.  By analogy, we wish to determine the
``climate'' of GRB afterglows, rather than simply looking at the ``weather''.
In general, we have a picture where the optical emission traces the forward
shock.  During the $\gamma$-ray emission, some afterglows show additional
components.  After the early time, the X-ray emission also appears to trace the
forward shock, but it is much more contaminated by prompt emission and flaring,
both of which can be attributed to activity of the central engine.  After
excluding X-ray flares, both the X-ray spectral index, $\betax$, and the
broadband optical-to-X-ray spectral index, $\betaox$, are consistent with the
fireball model predictions for the forward shock.  Thus, the optical and X-ray
emission are from the same emission region. In almost all cases, $\betax
\sim -1$ is a good description of the X-ray spectrum, and, after correcting for
extinction, the optical emission is consistent with a simple extrapolation of
the X-ray spectrum, or with the synchrotron cooling break, $\nu_c$, between the
optical and X-ray regime.

While the spectral indices are consistent across afterglows, and consistent
with the predictions of the fireball model, the decay indices are not.  We have
found the afterglows that fade rapidly in the optical also fade rapidly in the
X-rays.  The very slowly decaying GRB~060729 (Figure~\ref{fig:combo3}) and the
rapidly decaying GRB~061007 (Figure~\ref{fig:combo3}) are very different
temporally, yet have very similar spectral features, as illustrated in
Figure~\ref{fig:fnuopt_fnux}. In the case of the slowly decaying afterglows, we
can posit some form of long-duration energy injection from the central engine
to re-energize the blast wave, but this is somewhat ad-hoc, and does not
explain the break times~\citep{pmbng06}.  In addition, energy injection cannot
explain the rapidly decaying afterglows such as GRB~061007 and
GRB~050525a. \citet{mmkgg08} also noted that several afterglows are not
consistent with the fireball model, even after accounting for energy injection.
Although the absolute value of the decay index is incorrectly predicted by the
fireball model, the relative decay indices between the optical and X-rays is
roughly consistent with the predictions for a forward shock expanding into a
constant density medium. That is, for most afterglows, the optical decay is
slightly shallower than the X-ray decay or is consistent with the same
synchrotron regime.  The failure of the fireball model to correctly predict the
absolute value of the decay index, in light of the remarkably consistent X-ray
and broadband spectra, remains the fundamental limitation of the model.

\citet{woogp07} have suggested a model in which the X-ray afterglow can be
described as a simple combination of a rapidly decaying prompt component and
the rise and fall of the forward shock.  In broad strokes, this is consistent
with what we observe for most ROTSE-III afterglows.  Notably, the X-ray
afterglow of GRB~060605 (Figure~\ref{fig:combo2}) appears to a superposition of
the tail of the prompt emission and the optical afterglow, which traces the
onset of the forward shock.  Although this superposition model is able to
explain the flat portion of the X-ray afterglow without positing long-duration
energy injection, we observe a relatively steep optical and X-ray decay that is
significantly more rapid than predicted by using the X-ray spectral index and
standard fireball parameters.  Conversely, the extraordinarily long-lived
shallow decay of the afterglow of GRB~060729 also cannot be described by this
simple superposition model.  Thus, for many of the afterglows there remains an
inconsistency between the spectral and temporal properties.

In this paper we studied the relation between the early optical and X-ray
afterglow, focusing on the onset of the forward shock.  \citet{yaaab07} studied
the correlation (or lack thereof) between the prompt optical and $\gamma$-ray
emission for these and other ROTSE-III bursts.  They found that there is no
obvious correlation between the contemporaneous optical and $\gamma$-ray
emission at the earliest times.  As is the case with the optical and X-ray
comparisons in this paper, the vast majority of prompt optical detections and
limits are significantly dimmer than an extrapolation of the $\gamma$-ray
spectrum to the optical regime.  However, before the onset of the forward
shock, a small subset of prompt optical detections show excess
emission. GRB~990123~\citep{abbbb99,bbkpk99}, GRB~050820a~\citep{vwwag06},
GRB~051111~\citep{ysraa07,yaaab07}, GRB~061126~\citep{pbbph08}, and the
``naked-eye burst'' GRB~080319~\citep{rksgw08} all had optical flashes with
flux significantly in excess of an extrapolation of the contemporaneous
$\gamma$-ray emission.  In some cases (e.g., GRB~050820a, GRB~080319), the
optical flux was correlated with the $\gamma$-ray flux, and in other cases
(e.g., GRB~990123), they were not correlated.  The excess optical component may
be from a different emission region~\citep[e.g. large-angle
emission,][]{pk07,kp08}, or it may be dominated by a different emission
process~\citep[e.g. synchrotron and synchrotron self-Compton][]{kp08}.

For the 12 ROTSE-III bursts presented in this paper, none of the prompt optical
detections are obviously correlated with the prompt $\gamma$-ray emission.  For
example, optical afterglow of GRB~060111b has maximum flux prior to the second
$\gamma$-ray peak, and does not appear to be correlated.  On the other hand,
the single bright observation of GRB~060729 may coincide with one of the
$\gamma$-ray peaks, but not the other.  This was an exceptionally luminous GRB,
and the optical emission is significantly dimmer than an extrapolation of the
$\gamma$-ray spectrum.  Due to happenstance, none of the exceptionally bright
optical afterglows has been observed by ROTSE-III and XRT.  Instead, the
present prompt detections are more akin to GRB~061121~\citep{pwozg07} and
XRF~071031~\citep{kgmkr09}.  In the case of GRB~061121, the second $\gamma$-ray
peak is observed in the X-rays, NUV, and ROTSE-III band.  Notably, the peak is
very blue, and although the peak is prominent in the NUV, it is not
significantly observed in the $C_R$ filter.  Similarly, the high precision
photometry of the late afterglow of XRT~071031 shows that X-ray flares may be
visible in the optical, although they are also very blue in color.  Thus, the
red sensitivity of ROTSE-III may make it more difficult to see the contribution
from the prompt emission except for the most luminous optical flashes.

After the onset of self-similar evolution in the forward shock, we can use the
peak time of the forward shock to estimate the initial bulk Lorentz factor of
the outflow, $\Gamma_0$.  Through a broadband spectral analysis we have shown
that the optical afterglow is a relatively clean tracer of the forward external
shock, and thus we can use the optical peak time to estimate $\Gamma_0$.  For
the 12 bursts in this paper, this covers a wide range, from $\sim100$ for the
low luminosity, late peaking GRB~070611, to $\gtrsim900$ for the high
luminosity, early peaking GRB~050401.  For no bursts do we estimate
$\Gamma_0<100$, which has been estimated as the minimum for which the creation
of a non-thermal GRB spectrum is possible~\citep{p05}.  The range of Lorentz
factors is consistent with that observed in other
GRBs~\citep[e.g.,][]{sr02,mvmcd07}.  Although we have shown that the
fireball model is inadequate in predicting the absolute value of the temporal
indices of the early afterglow, the time of the onset and the implied
$\Gamma_0$ are both consistent with predictions~\citep[e.g.,][]{p05,gsw01}.

The \swift{} satellite has opened a new era of GRB observations, allowing
prompt and early multi-wavelength observations of a large sample of bursts and
afterglows.  By combining rapid GRB observations with the ROTSE-III telescope
network with XRT coverage, we have been able to peer into the fireball and
constrain the initial bulk Lorentz factor for a large number of GRBs.  Although
afterglows show common X-ray and broadband spectral properties that are
consistent with the predictions of the fireball model, the temporal properties
are quite different. It is not yet clear if a single model will be able to
explain the ``weather'' that determines the large variety of temporal decay
indices.

\acknowledgements

ESR would like to thank the TABASGO foundation.  This work has been supported
by NASA grant NNG-04WC41G, NSF grants AST-0407061 and PHY-0801007, the
Australian Research Council's \emph{Discovery Projects} funding scheme, the
University of New South Wales, the University of Texas, and the University of
Michigan.  HAF has been supported by NSF grant AST 03-35588 and by the Michigan
Space Grant Consortium.  FY has been supported under NASA \swift{} Guest
Investigator grants NNG-06GI90G and NNX-07AF02G.  JCW is supported in part
by NSF grant AST-0707769.  Special thanks to David Doss
at McDonald Observatory, Toni Hanke at the H.E.S.S. site, and Tuncay
\"{O}z{\i}\c{s}{\i}k at TUG.

\appendix

\section{ROTSE-III Photometry Tables}

\LongTables

\begin{deluxetable}{ccccccc}
\tablewidth{0pt}
\tablecaption{ROTSE-III $C_R$ Optical Photometry}
\tabletypesize{\scriptsize}
\tablehead{
\colhead{GRB} &
\colhead{Tel.} &
\colhead{$t_{\mathrm{start}}$ (s)} &
\colhead{$t_{\mathrm{end}}$ (s)} &
\colhead{$C_R$} &
\colhead{$f_{\nu,\mathrm{O}}$ (mJy)} &
\colhead{$f_{\nu,\mathrm{X}}$ (mJy)}
}
\label{tab:photom}
\startdata
050319 & IIIb &  164.1 &   169.1 & $15.97\pm 0.14$ & $1.48 \pm 0.19$ & \\
&&  178.5 &   183.5 & $16.31\pm 0.19$ & $1.09 \pm 0.19$ & \\
&&  192.9 &   197.9 & $16.18\pm 0.15$ & $1.22 \pm 0.16$ & \\
&&  207.5 &   212.5 & $16.31\pm 0.16$ & $1.09 \pm 0.16$ & \\
&&  222.1 &   227.1 & $16.86\pm 0.29$ & $0.652 \pm 0.171$ & \\
&&  236.4 &   241.4 & $16.22\pm 0.15$ & $1.17 \pm 0.16$ & \\
&&  250.9 &   270.5 & $16.67\pm 0.36$ & $0.781 \pm 0.262$ & $0.0253 \pm 0.0064$\\
&&  279.7 &   284.7 & $16.43\pm 0.18$ & $0.972 \pm 0.165$ & \\
&&  294.2 &   299.2 & $16.62\pm 0.24$ & $0.816 \pm 0.183$ & $0.00961 \pm 0.00363$\\
&&  308.5 &   328.5 & $16.89\pm 0.13$ & $0.635 \pm 0.079$ & $0.00630 \pm 0.00160$\\
&&  338.2 &   358.2 & $16.67\pm 0.10$ & $0.780 \pm 0.069$ & $0.00329 \pm 0.00109$\\
&&  367.7 &   387.7 & $16.79\pm 0.15$ & $0.700 \pm 0.100$ & $0.00269 \pm 0.00097$\\
&&  397.0 &   446.8 & $16.95\pm 0.15$ & $0.600 \pm 0.082$ & $0.00227 \pm 0.00060$\\
&&  456.0 &   476.0 & $16.96\pm 0.22$ & $0.596 \pm 0.119$ & $0.00118 \pm 0.00062$\\
&&  485.8 &   535.2 & $17.10\pm 0.14$ & $0.523 \pm 0.069$ & $0.00254 \pm 0.00065$\\
&&  544.7 &   564.7 & $17.05\pm 0.18$ & $0.548 \pm 0.093$ & $0.000582 \pm 0.000432$\\
&&  574.0 &   594.0 & $17.07\pm 0.20$ & $0.536 \pm 0.101$ & $0.00209 \pm 0.00084$\\
&&  603.8 &   663.8 & $17.44\pm 0.16$ & $0.382 \pm 0.056$ & $0.00108 \pm 0.00036$\\
&&  673.0 &   733.0 & $17.46\pm 0.20$ & $0.375 \pm 0.070$ & $0.00289 \pm 0.00066$\\
&&  742.2 &   802.2 & $17.45\pm 0.20$ & $0.379 \pm 0.071$ & $0.00179 \pm 0.00048$\\
&&  811.6 &   871.6 & $17.78\pm 0.24$ & $0.280 \pm 0.063$ & $0.00118 \pm 0.00038$\\
&&  881.2 &   941.2 & $17.42\pm 0.16$ & $0.389 \pm 0.056$ & $0.00118 \pm 0.00038$\\
&&  950.5 &  1080.2 & $18.00\pm 0.20$ & $0.228 \pm 0.041$ & $0.00147 \pm 0.00033$\\
&& 1089.4 &  1149.4 & $17.93\pm 0.28$ & $0.245 \pm 0.063$ & $0.00149 \pm 0.00043$\\
&& 1158.8 &  1218.8 & $18.11\pm 0.34$ & $0.207 \pm 0.065$ & $0.00108 \pm 0.00036$\\
&& 1228.6 &  1288.6 & $17.99\pm 0.31$ & $0.232 \pm 0.065$ & $0.00108 \pm 0.00036$\\
&& 1297.8 &  1427.1 & $18.29\pm 0.28$ & $0.176 \pm 0.045$ & $0.00105 \pm 0.00026$\\
&& 1437.0 &  1774.5 & $18.32\pm 0.21$ & $0.170 \pm 0.034$ & $0.00118 \pm 0.00021$\\
&& 1784.2 &  2052.9 & $18.29\pm 0.29$ & $0.174 \pm 0.046$ & \\
&& 2062.1 &  2746.9 & $19.50\pm 0.29$ & $0.0572 \pm 0.0152$ & \\
&& 2756.5 &  3163.6 & $18.84\pm 0.25$ & $0.106 \pm 0.024$ & \\
&& 3172.8 &  3719.5 & $18.70\pm 0.15$ & $0.120 \pm 0.017$ & \\
&& 3728.7 &  4345.9 & $18.72\pm 0.14$ & $0.117 \pm 0.015$ & \\
&& 4355.2 &  4902.0 & $18.67\pm 0.31$ & $0.124 \pm 0.035$ & \\
050401 & IIIa & 33.2 &    38.2 & $16.80\pm 0.29$ & $0.735 \pm 0.196$ & \\
&&   47.5 &    89.7 & $17.59\pm 0.34$ & $0.355 \pm 0.111$ & \\
&&   99.2 &   140.9 & $17.42\pm 0.23$ & $0.415 \pm 0.088$ & \\
&&  150.2 &   184.3 & $17.88\pm 0.25$ & $0.272 \pm 0.063$ & $0.0269 \pm 0.0018$\\
&&  201.5 &   281.2 & $18.58\pm 0.43$ & $0.143 \pm 0.056$ & $0.0208 \pm 0.0012$\\
050525a & IIIc & 363.4 &   449.6 & $15.35\pm 0.32$ & $2.81 \pm 0.83$ & $0.0610 \pm 0.0032$\\
&  & 458.4 &   567.2 & $14.97\pm 0.17$ & $4.00 \pm 0.63$ & $0.0510 \pm 0.0027$\\
&  & 577.1 &   715.0 & $15.93\pm 0.26$ & $1.66 \pm 0.39$ & $0.0473 \pm 0.0024$\\
&  & 724.5 &   922.5 & $15.80\pm 0.14$ & $1.87 \pm 0.23$ & $0.0396 \pm 0.0020$\\
&  & 932.2 &  1159.2 & $> 15.35$ & $>2.82$ & \\
& & 1168.5 &  1454.9 & $16.28\pm 0.22$ & $1.20 \pm 0.24$ & \\
& & 1464.9 &  1840.1 & $16.46\pm 0.07$ & $1.01 \pm 0.07$ & \\
& & 1849.8 &  2316.2 & $16.43\pm 0.07$ & $1.04 \pm 0.07$ & \\
& & 2326.1 &  2908.2 & $16.44\pm 0.06$ & $1.03 \pm 0.06$ & \\
& & 2917.6 &  3647.0 & $16.61\pm 0.07$ & $0.887 \pm 0.053$ & \\
& & 3656.2 &  4593.1 & $16.67\pm 0.07$ & $0.836 \pm 0.050$ & \\
& & 4602.4 &  5804.2 & $16.97\pm 0.08$ & $0.632 \pm 0.046$ & \\
& & 5813.9 &  7310.9 & $17.30\pm 0.09$ & $0.466 \pm 0.037$ & $0.00135 \pm 0.00014$\\ 
& & 7320.0 &  9203.4 & $17.56\pm 0.18$ & $0.367 \pm 0.060$ & \\
& & 9213.0 & 12513.8 & $17.74\pm 0.15$ & $0.313 \pm 0.044$ & $0.000588 \pm 0.000070$\\
& IIId & 2348.1 &  2496.3 & $16.56\pm 0.13$ & $0.928 \pm 0.112$ & \\
& & 2513.4 &  2707.6 & $16.43\pm 0.06$ & $1.04 \pm 0.06$ & \\
& & 2717.6 &  2883.1 & $16.47\pm 0.08$ & $1.00 \pm 0.08$ & \\
& & 2891.9 &  3085.2 & $16.51\pm 0.07$ & $0.973 \pm 0.065$ & \\
& & 3094.1 &  3316.4 & $16.37\pm 0.07$ & $1.10 \pm 0.07$ & \\
& & 3325.2 &  3548.3 & $16.40\pm 0.07$ & $1.07 \pm 0.06$ & \\
& & 3557.1 &  3809.4 & $16.57\pm 0.07$ & $0.918 \pm 0.060$ & \\
& & 3819.1 &  4071.4 & $16.66\pm 0.08$ & $0.842 \pm 0.063$ & \\
& & 4080.0 &  4487.9 & $16.67\pm 0.11$ & $0.835 \pm 0.083$ & \\
050801 & IIIc &    21.8 &    26.8 & $14.93\pm 0.05$ & $4.16 \pm 0.17$ & \\
&&   29.9 &    34.9 & $14.79\pm 0.05$ & $4.74 \pm 0.20$ & \\
&&   38.0 &    43.0 & $14.80\pm 0.04$ & $4.71 \pm 0.19$ & \\
&&   46.1 &    51.1 & $14.91\pm 0.06$ & $4.24 \pm 0.23$ & \\
&&   54.2 &    59.2 & $14.83\pm 0.05$ & $4.54 \pm 0.22$ & \\
&&   62.4 &    67.4 & $14.91\pm 0.04$ & $4.25 \pm 0.18$ & \\
&&   70.5 &    75.5 & $14.75\pm 0.04$ & $4.90 \pm 0.17$ & $0.00877 \pm 0.00561$\\
&&   78.6 &    83.6 & $14.87\pm 0.05$ & $4.41 \pm 0.20$ & $0.00491 \pm 0.00343$\\
&&   86.7 &    91.7 & $14.88\pm 0.05$ & $4.37 \pm 0.22$ & \\
&&   94.8 &    99.8 & $14.93\pm 0.05$ & $4.15 \pm 0.19$ & $0.00361 \pm 0.00193$\\
&&  113.5 &   133.5 & $14.98\pm 0.03$ & $3.96 \pm 0.11$ & $0.00338 \pm 0.00107$\\
&&  143.3 &   163.3 & $15.09\pm 0.03$ & $3.58 \pm 0.09$ & $0.00202 \pm 0.00077$\\
&&  172.7 &   192.7 & $15.12\pm 0.03$ & $3.50 \pm 0.09$ & $0.00293 \pm 0.00097$\\
&&  203.0 &   223.0 & $15.06\pm 0.03$ & $3.69 \pm 0.10$ & $0.00429 \pm 0.00125$\\
&&  232.5 &   252.5 & $15.13\pm 0.04$ & $3.48 \pm 0.13$ & $0.00338 \pm 0.00107$\\
&&  262.3 &   282.3 & $15.21\pm 0.04$ & $3.21 \pm 0.11$ & $0.00180 \pm 0.00072$\\
&&  291.8 &   311.8 & $15.35\pm 0.04$ & $2.83 \pm 0.10$ & $0.00338 \pm 0.00107$\\
&&  321.0 &   341.0 & $15.47\pm 0.04$ & $2.54 \pm 0.09$ & $0.00338 \pm 0.00107$\\
&&  350.8 &   370.8 & $15.59\pm 0.03$ & $2.27 \pm 0.07$ & $0.00225 \pm 0.00082$\\
&&  380.3 &   400.3 & $15.70\pm 0.04$ & $2.04 \pm 0.08$ & $0.00270 \pm 0.00092$\\
&&  409.9 &   469.9 & $15.89\pm 0.04$ & $1.71 \pm 0.07$ & $0.00172 \pm 0.00048$\\
&&  479.8 &   539.8 & $16.12\pm 0.03$ & $1.39 \pm 0.04$ & $0.00142 \pm 0.00042$\\
&&  549.0 &   609.0 & $16.29\pm 0.04$ & $1.19 \pm 0.04$ & $0.00119 \pm 0.00037$\\
&&  618.2 &   678.2 & $16.31\pm 0.05$ & $1.16 \pm 0.06$ & $0.000586 \pm 0.000238$\\
&&  688.1 &   748.1 & $16.63\pm 0.06$ & $0.873 \pm 0.050$ & $0.000662 \pm 0.000256$\\
&&  757.2 &   817.2 & $16.59\pm 0.06$ & $0.901 \pm 0.051$ & $0.000813 \pm 0.000290$\\
&&  826.6 &   886.6 & $16.66\pm 0.07$ & $0.844 \pm 0.056$ & $0.000586 \pm 0.000238$\\
&&  896.3 &   956.3 & $16.75\pm 0.06$ & $0.780 \pm 0.044$ & $0.000737 \pm 0.000273$\\
&&  965.5 &  1025.5 & $16.93\pm 0.07$ & $0.658 \pm 0.045$ & $0.00232 \pm 0.00171$\\
&& 1034.9 &  1094.9 & $16.92\pm 0.09$ & $0.665 \pm 0.055$ & \\
&& 1104.7 &  1233.9 & $16.99\pm 0.06$ & $0.624 \pm 0.032$ & \\
&& 1243.6 &  1442.0 & $17.10\pm 0.05$ & $0.563 \pm 0.024$ & \\
&& 1451.4 &  1650.3 & $17.39\pm 0.07$ & $0.432 \pm 0.027$ & \\
&& 1659.7 &  1858.6 & $17.48\pm 0.07$ & $0.397 \pm 0.025$ & \\
&& 1867.9 &  2136.8 & $17.60\pm 0.06$ & $0.355 \pm 0.021$ & \\
&& 2146.5 &  2485.3 & $17.78\pm 0.07$ & $0.303 \pm 0.019$ & \\
&& 2495.2 &  2832.6 & $17.88\pm 0.07$ & $0.274 \pm 0.018$ & \\
&& 2841.9 &  3249.7 & $18.26\pm 0.11$ & $0.194 \pm 0.020$ & \\
&& 3259.7 &  3736.8 & $18.24\pm 0.09$ & $0.198 \pm 0.017$ & \\
&& 3745.9 &  4332.1 & $18.71\pm 0.20$ & $0.128 \pm 0.023$ & \\
&& 4341.4 &  4956.6 & $18.49\pm 0.09$ & $0.157 \pm 0.013$ & $0.000177 \pm 0.000047$\\
&& 4966.5 &  5721.7 & $18.88\pm 0.12$ & $0.109 \pm 0.012$ & $0.0000900 \pm 0.0000293$\\
&& 5731.0 &  6554.7 & $18.99\pm 0.15$ & $0.0993 \pm 0.0135$ & $0.0000673 \pm 0.0000208$\\
&& 6564.4 &  7527.4 & $18.83\pm 0.13$ & $0.115 \pm 0.014$ & \\
&& 7536.7 &  8619.8 & $19.63\pm 0.22$ & $0.0549 \pm 0.0110$ & \\
&& 8629.6 & 10357.0 & $19.49\pm 0.16$ & $0.0625 \pm 0.0092$ & \\
050922c & IIId & 172.4 &   177.4 & $14.58\pm 0.07$ & $5.92 \pm 0.40$ & $0.0224 \pm 0.0033$\\
&&  186.8 &   191.8 & $14.62\pm 0.08$ & $5.67 \pm 0.41$ & $0.0183 \pm 0.0029$\\
&&  200.6 &   205.6 & $14.64\pm 0.08$ & $5.55 \pm 0.39$ & $0.0253 \pm 0.0035$\\
&&  214.4 &   219.4 & $14.76\pm 0.07$ & $5.00 \pm 0.34$ & $0.0212 \pm 0.0032$\\
&&  228.2 &   233.2 & $14.81\pm 0.08$ & $4.78 \pm 0.36$ & $0.0171 \pm 0.0028$\\
&&  242.2 &   247.2 & $14.99\pm 0.09$ & $4.03 \pm 0.32$ & $0.0157 \pm 0.0027$\\
&&  255.9 &   260.9 & $15.03\pm 0.09$ & $3.88 \pm 0.34$ & $0.0157 \pm 0.0027$\\
&&  269.9 &   274.9 & $15.13\pm 0.09$ & $3.55 \pm 0.30$ & $0.0190 \pm 0.0029$\\
&&  284.2 &   289.2 & $15.03\pm 0.10$ & $3.90 \pm 0.34$ & $0.0160 \pm 0.0027$\\
&&  298.6 &   303.6 & $15.18\pm 0.09$ & $3.40 \pm 0.30$ & $0.0160 \pm 0.0027$\\
&&  312.3 &   332.3 & $15.23\pm 0.06$ & $3.22 \pm 0.17$ & $0.0143 \pm 0.0015$\\
&&  341.3 &   361.3 & $15.37\pm 0.05$ & $2.84 \pm 0.12$ & $0.0106 \pm 0.0012$\\
&&  370.2 &   390.2 & $15.48\pm 0.07$ & $2.56 \pm 0.18$ & $0.0113 \pm 0.0013$\\
&&  399.2 &   419.2 & $15.49\pm 0.07$ & $2.55 \pm 0.16$ & $0.0102 \pm 0.0012$\\
&&  428.0 &   448.0 & $15.57\pm 0.06$ & $2.38 \pm 0.14$ & $0.00783 \pm 0.00148$\\
&&  457.2 &   477.2 & $15.54\pm 0.06$ & $2.44 \pm 0.13$ & $0.00736 \pm 0.00197$\\
&&  486.0 &   506.0 & $15.67\pm 0.08$ & $2.17 \pm 0.16$ & $0.00461 \pm 0.00149$\\
&&  514.9 &   534.9 & $15.68\pm 0.08$ & $2.14 \pm 0.17$ & $0.00854 \pm 0.00216$\\
&&  543.8 &   563.8 & $15.82\pm 0.08$ & $1.89 \pm 0.13$ & $0.00658 \pm 0.00184$\\
&&  573.0 &   593.0 & $15.76\pm 0.10$ & $1.99 \pm 0.18$ & $0.00697 \pm 0.00191$\\
&&  601.7 &   621.7 & $15.84\pm 0.08$ & $1.85 \pm 0.13$ & $0.00894 \pm 0.00223$\\
&&  630.6 &   650.6 & $15.85\pm 0.08$ & $1.83 \pm 0.14$ & $0.00540 \pm 0.00164$\\
&& 3005.3 &  3198.6 & $17.05\pm 0.09$ & $0.605 \pm 0.052$ & \\
&& 3207.7 &  3400.8 & $16.98\pm 0.06$ & $0.643 \pm 0.038$ & \\
&& 3409.4 &  3851.5 & $17.12\pm 0.04$ & $0.567 \pm 0.020$ & \\
051109a & IIIb & 35.4 &    40.4 & $14.99\pm 0.06$ & $5.03 \pm 0.28$ & \\
&&   42.5 &    47.5 & $15.00\pm 0.06$ & $4.99 \pm 0.29$ & \\
&&   49.6 &    54.6 & $15.15\pm 0.07$ & $4.34 \pm 0.28$ & \\
&&   56.7 &    61.7 & $15.20\pm 0.07$ & $4.15 \pm 0.27$ & \\
&&   63.8 &    68.8 & $15.35\pm 0.08$ & $3.62 \pm 0.27$ & \\
&&   70.9 &    75.9 & $15.31\pm 0.08$ & $3.76 \pm 0.27$ & \\
&&   78.0 &    83.0 & $15.44\pm 0.09$ & $3.31 \pm 0.27$ & \\
&&   85.1 &    90.1 & $15.48\pm 0.09$ & $3.21 \pm 0.27$ & \\
&&   92.3 &    97.3 & $15.37\pm 0.08$ & $3.55 \pm 0.25$ & \\
&&   99.4 &   104.4 & $15.53\pm 0.09$ & $3.06 \pm 0.26$ & \\
&&  117.7 &   137.7 & $15.70\pm 0.05$ & $2.61 \pm 0.13$ & $0.0404 \pm 0.0053$\\
&&  154.8 &   174.8 & $15.90\pm 0.06$ & $2.18 \pm 0.12$ & $0.0208 \pm 0.0027$\\
&&  184.4 &   204.4 & $15.96\pm 0.07$ & $2.06 \pm 0.13$ & $0.0111 \pm 0.0016$\\
&&  213.3 &   233.3 & $15.92\pm 0.06$ & $2.14 \pm 0.12$ & \\
&&  243.6 &   263.6 & $16.08\pm 0.07$ & $1.84 \pm 0.12$ & \\
&&  272.9 &   292.9 & $16.21\pm 0.08$ & $1.64 \pm 0.12$ & \\
&&  302.0 &   322.0 & $16.20\pm 0.07$ & $1.65 \pm 0.11$ & \\
&&  331.0 &   351.0 & $16.48\pm 0.10$ & $1.28 \pm 0.11$ & \\
&&  360.1 &   380.1 & $16.55\pm 0.11$ & $1.19 \pm 0.12$ & \\
&&  389.7 &   409.7 & $16.50\pm 0.10$ & $1.25 \pm 0.11$ & \\
&&  419.2 &   479.2 & $16.47\pm 0.06$ & $1.29 \pm 0.07$ & \\
&&  488.4 &   548.4 & $16.75\pm 0.07$ & $0.992 \pm 0.064$ & \\
&&  557.5 &   617.5 & $16.81\pm 0.07$ & $0.945 \pm 0.064$ & \\
&&  626.7 &   686.7 & $16.99\pm 0.09$ & $0.800 \pm 0.069$ & \\
&&  695.8 &   755.8 & $16.91\pm 0.08$ & $0.857 \pm 0.066$ & \\
&&  764.8 &   824.8 & $17.06\pm 0.10$ & $0.750 \pm 0.069$ & \\
&&  833.9 &   893.9 & $17.06\pm 0.10$ & $0.745 \pm 0.067$ & \\
&&  903.1 &   963.1 & $17.37\pm 0.14$ & $0.563 \pm 0.070$ & \\
&&  972.2 &  1032.2 & $17.59\pm 0.16$ & $0.459 \pm 0.067$ & \\
&& 1041.4 &  1101.4 & $17.19\pm 0.11$ & $0.664 \pm 0.068$ & \\
&& 1111.2 &  1448.6 & $17.41\pm 0.07$ & $0.541 \pm 0.034$ & \\
&& 1457.5 &  1794.0 & $17.56\pm 0.08$ & $0.472 \pm 0.035$ & \\
&& 1803.3 &  2139.8 & $17.86\pm 0.10$ & $0.357 \pm 0.031$ & \\
&& 2148.9 &  3177.3 & $17.89\pm 0.08$ & $0.348 \pm 0.025$ & \\
&& 3186.5 &  4366.5 & $18.29\pm 0.14$ & $0.242 \pm 0.030$ & $0.00243 \pm 0.00027$\\
&& 4375.8 &  5403.9 & $18.19\pm 0.11$ & $0.264 \pm 0.026$ & $0.00204 \pm 0.00023$\\
&& 5413.0 &  7477.9 & $18.31\pm 0.11$ & $0.236 \pm 0.024$ & \\
&& 7487.9 &  9556.3 & $18.46\pm 0.12$ & $0.206 \pm 0.022$ & \\
&& 9565.2 & 12045.7 & $18.57\pm 0.13$ & $0.187 \pm 0.022$ & $0.000922 \pm 0.000103$\\
&&12055.0 & 14534.9 & $18.90\pm 0.17$ & $0.137 \pm 0.021$ & \\
060111b & IIId & 32.8 &    37.8 & $13.11\pm 0.05$ & $23.0 \pm  1.1$ & \\
&&   46.0 &    51.0 & $14.12\pm 0.09$ & $9.12 \pm 0.74$ & \\
&&   59.7 &    64.7 & $14.56\pm 0.12$ & $6.05 \pm 0.68$ & \\
&&   73.3 &    78.3 & $15.23\pm 0.17$ & $3.27 \pm 0.51$ & \\
&&   87.2 &    92.2 & $15.62\pm 0.27$ & $2.29 \pm 0.56$ & $0.0538 \pm 0.0162$\\
&&  100.8 &   105.8 & $15.83\pm 0.19$ & $1.89 \pm 0.34$ & $0.0360 \pm 0.0071$\\
&&  114.3 &   133.2 & $16.54\pm 0.35$ & $0.977 \pm 0.311$ & $0.0148 \pm 0.0027$\\
&&  141.7 &   160.1 & $16.62\pm 0.28$ & $0.909 \pm 0.236$ & $0.00734 \pm 0.00194$\\
&&  168.7 &   188.7 & $16.90\pm 0.23$ & $0.702 \pm 0.146$ & $0.00749 \pm 0.00184$\\
&&  197.0 &   245.2 & $> 16.73$ & $>0.825$ & \\
&&  451.9 &   728.6 & $> 17.72$ & $>0.330$ & \\
060605 & IIIa &    49.4 &    54.4 & $> 16.32$ & $>1.46$ & \\
&&   71.4 &    76.4 & $16.51\pm 0.29$ & $1.23 \pm 0.33$ & \\
&&   94.0 &    99.0 & $16.09\pm 0.24$ & $1.81 \pm 0.40$ & $0.00559 \pm 0.00573$\\
&&  116.4 &   121.4 & $15.89\pm 0.15$ & $2.17 \pm 0.31$ & $0.00802 \pm 0.00264$\\
&&  138.3 &   143.3 & $15.87\pm 0.17$ & $2.22 \pm 0.36$ & $0.00443 \pm 0.00261$\\
&&  160.2 &   165.2 & $15.48\pm 0.11$ & $3.17 \pm 0.32$ & $0.00295 \pm 0.00212$\\
&&  182.3 &   187.3 & $15.62\pm 0.16$ & $2.78 \pm 0.42$ & $0.00295 \pm 0.00212$\\
&&  204.6 &   209.6 & $15.46\pm 0.16$ & $3.22 \pm 0.48$ & $0.00591 \pm 0.00302$\\
&&  226.9 &   231.9 & $15.47\pm 0.14$ & $3.19 \pm 0.41$ & $0.00146 \pm 0.00149$\\
&&  240.7 &   245.7 & $15.71\pm 0.18$ & $2.55 \pm 0.42$ & $0.00146 \pm 0.00149$\\
&&  263.0 &   283.0 & $15.47\pm 0.06$ & $3.21 \pm 0.18$ & $0.00295 \pm 0.00109$\\
&&  300.3 &   320.3 & $15.45\pm 0.07$ & $3.27 \pm 0.21$ & $0.00332 \pm 0.00116$\\
&&  337.6 &   357.6 & $15.54\pm 0.09$ & $2.99 \pm 0.24$ & $0.00146 \pm 0.00076$\\
&&  374.5 &   394.5 & $15.56\pm 0.08$ & $2.95 \pm 0.22$ & $0.00146 \pm 0.00076$\\
&&  411.7 &   431.7 & $15.33\pm 0.08$ & $3.64 \pm 0.26$ & $0.00221 \pm 0.00093$\\
&&  449.1 &   469.1 & $15.35\pm 0.09$ & $3.56 \pm 0.29$ & $0.000351 \pm 0.000373$\\
&&  486.7 &   506.7 & $15.42\pm 0.08$ & $3.34 \pm 0.23$ & $0.000351 \pm 0.000373$\\
&&  523.9 &   543.9 & $15.40\pm 0.05$ & $3.42 \pm 0.17$ & \\
&&  560.9 &   580.9 & $15.41\pm 0.10$ & $3.39 \pm 0.30$ & \\
&&  598.1 &   618.1 & $15.37\pm 0.04$ & $3.52 \pm 0.14$ & \\
&&  635.5 &   695.5 & $15.54\pm 0.07$ & $3.00 \pm 0.19$ & \\
&&  713.1 &   773.1 & $15.46\pm 0.07$ & $3.22 \pm 0.22$ & \\
&&  790.0 &   850.0 & $15.58\pm 0.05$ & $2.89 \pm 0.15$ & \\
&&  867.6 &   927.6 & $15.74\pm 0.08$ & $2.50 \pm 0.17$ & \\
&&  936.4 &   996.4 & $15.72\pm 0.06$ & $2.53 \pm 0.15$ & \\
&& 1005.6 &  1065.6 & $15.83\pm 0.08$ & $2.30 \pm 0.17$ & \\
&& 1082.6 &  1142.6 & $15.88\pm 0.08$ & $2.19 \pm 0.17$ & \\
&& 1159.5 &  1219.5 & $15.98\pm 0.08$ & $2.00 \pm 0.15$ & \\
&& 1236.7 &  1296.7 & $16.13\pm 0.09$ & $1.75 \pm 0.15$ & \\
&& 1305.8 &  1365.8 & $16.35\pm 0.09$ & $1.43 \pm 0.12$ & \\
&& 1383.4 &  1443.4 & $16.18\pm 0.06$ & $1.66 \pm 0.08$ & \\
&& 1452.9 &  1512.9 & $16.22\pm 0.08$ & $1.60 \pm 0.13$ & \\
&& 1522.3 &  1582.3 & $16.23\pm 0.10$ & $1.59 \pm 0.15$ & \\
&& 1591.6 &  1651.6 & $16.39\pm 0.18$ & $1.37 \pm 0.22$ & \\
&& 1660.5 &  1720.5 & $16.26\pm 0.10$ & $1.55 \pm 0.15$ & \\
&& 1730.1 &  1790.1 & $16.54\pm 0.10$ & $1.19 \pm 0.11$ & \\
&& 1799.7 &  1859.7 & $16.47\pm 0.12$ & $1.27 \pm 0.14$ & \\
&& 1869.2 &  1929.2 & $16.57\pm 0.13$ & $1.16 \pm 0.14$ & \\
&& 1938.4 &  1998.4 & $16.60\pm 0.15$ & $1.13 \pm 0.16$ & \\
&& 2007.4 &  2067.4 & $16.69\pm 0.17$ & $1.04 \pm 0.16$ & \\
&& 2077.0 &  2137.0 & $16.57\pm 0.19$ & $1.16 \pm 0.20$ & \\
&& 2146.2 &  2206.2 & $16.72\pm 0.17$ & $1.01 \pm 0.16$ & \\
&& 2215.6 &  2275.6 & $16.46\pm 0.09$ & $1.28 \pm 0.10$ & \\
&& 2285.0 &  2345.0 & $16.38\pm 0.08$ & $1.39 \pm 0.11$ & \\
&& 2354.0 &  2414.0 & $16.77\pm 0.13$ & $0.963 \pm 0.114$ & \\
&& 2422.9 &  2482.9 & $16.68\pm 0.14$ & $1.05 \pm 0.13$ & \\
&& 2492.5 &  2552.5 & $16.67\pm 0.13$ & $1.06 \pm 0.12$ & \\
&& 2561.5 &  2621.5 & $16.82\pm 0.18$ & $0.921 \pm 0.152$ & \\
&& 2630.3 &  2690.3 & $16.99\pm 0.18$ & $0.792 \pm 0.130$ & \\
&& 2699.5 &  2759.5 & $16.80\pm 0.16$ & $0.943 \pm 0.137$ & \\
&& 2768.6 &  3451.6 & $16.97\pm 0.06$ & $0.804 \pm 0.041$ & \\
&& 3460.9 &  4144.3 & $17.23\pm 0.06$ & $0.630 \pm 0.036$ & $0.000976 \pm 0.000170$\\
&& 4486.6 &  5200.6 & $17.72\pm 0.11$ & $0.403 \pm 0.042$ & $0.000739 \pm 0.000115$\\
&& 5210.0 &  5948.0 & $17.90\pm 0.13$ & $0.341 \pm 0.042$ & $0.000785 \pm 0.000118$\\
&& 5956.8 &  6677.6 & $17.72\pm 0.14$ & $0.403 \pm 0.050$ & $0.000656 \pm 0.000128$\\
060729 & IIIa &    64.5 &    69.5 & $> 16.70$ & $>0.738$ & \\
&&   78.6 &    83.6 & $16.99\pm 0.25$ & $0.562 \pm 0.132$ & \\
&&   93.1 &    98.1 & $15.79\pm 0.14$ & $1.70 \pm 0.21$ & \\
&&  115.0 &   120.0 & $17.07\pm 0.38$ & $0.523 \pm 0.184$ & \\
&&  129.0 &   134.0 & $16.96\pm 0.26$ & $0.578 \pm 0.136$ & $2.21 \pm 0.29$\\
&&  151.3 &   156.3 & $17.71\pm 0.30$ & $0.290 \pm 0.079$ & $0.793 \pm 0.107$\\
&&  165.2 &   170.2 & $17.21\pm 0.21$ & $0.459 \pm 0.089$ & $0.664 \pm 0.091$\\
&&  187.3 &   192.3 & $> 16.60$ & $>0.808$ & \\
&&  201.2 &   206.2 & $> 16.68$ & $>0.750$ & \\
&&  215.2 &   220.2 & $17.44\pm 0.52$ & $0.372 \pm 0.179$ & $0.210 \pm 0.031$\\
&&  229.8 &   249.8 & $17.41\pm 0.20$ & $0.383 \pm 0.071$ & $0.119 \pm 0.017$\\
&&  258.7 &   278.7 & $17.40\pm 0.24$ & $0.384 \pm 0.086$ & $0.0503 \pm 0.0076$\\
&&  295.6 &   315.6 & $17.08\pm 0.16$ & $0.520 \pm 0.075$ & $0.0189 \pm 0.0034$\\
&&  324.6 &   344.6 & $16.80\pm 0.10$ & $0.669 \pm 0.065$ & $0.00943 \pm 0.00211$\\
&&  354.0 &   374.0 & $17.02\pm 0.15$ & $0.550 \pm 0.074$ & $0.00426 \pm 0.00080$\\
&&  382.9 &   402.9 & $16.78\pm 0.14$ & $0.681 \pm 0.088$ & $0.00234 \pm 0.00055$\\
&&  419.8 &   439.8 & $16.77\pm 0.09$ & $0.692 \pm 0.057$ & $0.00283 \pm 0.00061$\\
&&  448.8 &   468.8 & $16.66\pm 0.10$ & $0.762 \pm 0.067$ & $0.00308 \pm 0.00063$\\
&&  478.1 &   498.1 & $16.74\pm 0.15$ & $0.711 \pm 0.100$ & $0.00209 \pm 0.00052$\\
&&  515.7 &   535.7 & $16.91\pm 0.14$ & $0.609 \pm 0.076$ & $0.00184 \pm 0.00049$\\
&&  544.9 &   604.9 & $16.59\pm 0.10$ & $0.816 \pm 0.079$ & $0.00151 \pm 0.00026$\\
&&  622.4 &   682.4 & $16.70\pm 0.14$ & $0.737 \pm 0.093$ & $0.00122 \pm 0.00023$\\
&&  691.8 &   751.8 & $16.89\pm 0.05$ & $0.615 \pm 0.029$ & $0.00110 \pm 0.00022$\\
&&  769.2 &   829.2 & $16.87\pm 0.14$ & $0.631 \pm 0.080$ & $0.00106 \pm 0.00022$\\
&&  838.0 &   898.0 & $16.71\pm 0.11$ & $0.731 \pm 0.075$ & $0.000977 \pm 0.000207$\\
&&  907.2 &   967.2 & $16.80\pm 0.09$ & $0.673 \pm 0.057$ & $0.00127 \pm 0.00024$\\
&&  984.3 &  1044.3 & $16.96\pm 0.11$ & $0.578 \pm 0.060$ & $0.000977 \pm 0.000207$\\
&& 1053.6 &  1113.6 & $16.80\pm 0.08$ & $0.673 \pm 0.052$ & $0.000812 \pm 0.000188$\\
&& 1122.9 &  1182.9 & $16.83\pm 0.09$ & $0.651 \pm 0.053$ & $0.00102 \pm 0.00021$\\
&& 1200.3 &  1260.3 & $16.91\pm 0.09$ & $0.607 \pm 0.049$ & $0.00102 \pm 0.00021$\\
&& 1269.6 &  1329.6 & $17.04\pm 0.10$ & $0.536 \pm 0.049$ & $0.000977 \pm 0.000207$\\
&& 1338.9 &  1398.9 & $16.86\pm 0.13$ & $0.635 \pm 0.075$ & $0.000853 \pm 0.000193$\\
&& 1415.9 &  1475.9 & $16.86\pm 0.12$ & $0.633 \pm 0.069$ & $0.00118 \pm 0.00023$\\
&& 1485.0 &  1545.0 & $16.86\pm 0.11$ & $0.636 \pm 0.062$ & $0.000977 \pm 0.000207$\\
&& 1553.9 &  1613.9 & $16.94\pm 0.10$ & $0.591 \pm 0.056$ & $0.00110 \pm 0.00022$\\
&& 1631.1 &  1691.1 & $16.90\pm 0.10$ & $0.612 \pm 0.057$ & $0.000729 \pm 0.000178$\\
&& 1700.4 &  1760.4 & $16.96\pm 0.11$ & $0.580 \pm 0.056$ & $0.00139 \pm 0.00025$\\
&& 1777.4 &  1837.4 & $16.98\pm 0.10$ & $0.571 \pm 0.051$ & $0.000977 \pm 0.000207$\\
&& 1846.6 &  1906.6 & $17.18\pm 0.11$ & $0.475 \pm 0.050$ & $0.00118 \pm 0.00023$\\
&& 1916.0 &  1976.0 & $16.93\pm 0.12$ & $0.595 \pm 0.067$ & $0.00119 \pm 0.00031$\\
&& 1985.4 &  2045.4 & $16.87\pm 0.08$ & $0.630 \pm 0.048$ & \\
&& 2054.5 &  2114.5 & $16.96\pm 0.12$ & $0.580 \pm 0.065$ & \\
&& 2123.7 &  2183.7 & $17.08\pm 0.08$ & $0.520 \pm 0.038$ & \\
&& 2201.2 &  2261.2 & $17.10\pm 0.15$ & $0.508 \pm 0.072$ & \\
&& 2270.1 &  2330.1 & $17.06\pm 0.13$ & $0.526 \pm 0.063$ & \\
&& 2339.2 &  2399.2 & $17.16\pm 0.12$ & $0.480 \pm 0.055$ & \\
&& 2416.3 &  2476.3 & $16.83\pm 0.12$ & $0.655 \pm 0.069$ & \\
&& 2485.5 &  2545.5 & $17.14\pm 0.18$ & $0.491 \pm 0.084$ & \\
&& 2554.7 &  2614.7 & $17.59\pm 0.16$ & $0.323 \pm 0.047$ & \\
&& 2631.5 &  2691.5 & $17.14\pm 0.16$ & $0.488 \pm 0.071$ & \\
&& 2701.0 &  2761.0 & $16.97\pm 0.10$ & $0.572 \pm 0.055$ & \\
&& 2770.3 &  2830.3 & $17.04\pm 0.20$ & $0.538 \pm 0.100$ & \\
&& 2847.1 &  2907.1 & $17.22\pm 0.30$ & $0.457 \pm 0.128$ & \\
&& 2916.6 &  2976.6 & $17.19\pm 0.38$ & $0.468 \pm 0.166$ & \\
&& 2985.9 &  3045.9 & $16.80\pm 0.28$ & $0.670 \pm 0.176$ & \\
060904b & IIIc &    19.3 &    24.3 & $> 16.91$ & $>0.810$ & \\
&&   27.4 &    32.4 & $17.45\pm 0.40$ & $0.495 \pm 0.183$ & \\
&&   35.4 &    40.4 & $17.09\pm 0.26$ & $0.691 \pm 0.164$ & \\
&&   43.4 &    48.4 & $> 16.84$ & $>0.866$ & \\
&&   51.4 &    56.4 & $16.78\pm 0.27$ & $0.919 \pm 0.231$ & \\
&&   59.5 &    64.5 & $> 16.95$ & $>0.782$ & \\
&&   67.5 &    72.5 & $17.03\pm 0.34$ & $0.725 \pm 0.228$ & \\
&&   75.5 &    80.5 & $17.13\pm 0.32$ & $0.666 \pm 0.198$ & $0.0157 \pm 0.0056$\\
&&   83.5 &    88.5 & $17.41\pm 0.39$ & $0.513 \pm 0.186$ & $0.0140 \pm 0.0044$\\
&&   91.5 &    96.5 & $17.40\pm 0.39$ & $0.519 \pm 0.187$ & \\
&&  110.1 &   130.1 & $17.92\pm 0.45$ & $0.320 \pm 0.132$ & $0.00906 \pm 0.00200$\\
&&  147.3 &   167.3 & $17.32\pm 0.21$ & $0.555 \pm 0.109$ & $0.705 \pm 0.033$\\
&&  177.5 &   197.5 & $17.14\pm 0.24$ & $0.657 \pm 0.143$ & $0.802 \pm 0.038$\\
&&  207.1 &   227.1 & $17.75\pm 0.33$ & $0.375 \pm 0.114$ & $0.390 \pm 0.020$\\
&&  237.0 &   257.0 & $17.13\pm 0.17$ & $0.664 \pm 0.106$ & $0.195 \pm 0.011$\\
&&  266.8 &   286.8 & $17.45\pm 0.21$ & $0.495 \pm 0.097$ & $0.0932 \pm 0.0064$\\
&&  296.3 &   316.3 & $16.67\pm 0.11$ & $1.02 \pm 0.10$ & \\
&&  325.4 &   345.4 & $16.92\pm 0.14$ & $0.804 \pm 0.103$ & \\
&&  354.8 &   374.8 & $16.72\pm 0.14$ & $0.972 \pm 0.126$ & \\
&&  384.3 &   404.3 & $16.65\pm 0.15$ & $1.03 \pm 0.14$ & \\
&&  413.9 &   473.9 & $16.68\pm 0.11$ & $1.01 \pm 0.10$ & \\
&&  483.2 &   543.2 & $16.63\pm 0.08$ & $1.05 \pm 0.08$ & \\
&&  553.0 &   613.0 & $16.52\pm 0.10$ & $1.16 \pm 0.10$ & \\
&&  622.1 &   682.1 & $16.92\pm 0.13$ & $0.801 \pm 0.093$ & \\
&&  691.5 &   751.5 & $16.86\pm 0.11$ & $0.847 \pm 0.083$ & \\
&&  761.2 &   821.2 & $17.06\pm 0.12$ & $0.709 \pm 0.080$ & \\
&&  830.4 &   890.4 & $17.45\pm 0.18$ & $0.495 \pm 0.083$ & \\
&&  899.7 &   959.7 & $17.15\pm 0.15$ & $0.654 \pm 0.089$ & \\
&&  969.5 &  1029.5 & $17.38\pm 0.17$ & $0.525 \pm 0.081$ & \\
&& 1038.8 &  1098.8 & $17.53\pm 0.18$ & $0.459 \pm 0.074$ & \\
&& 1108.0 &  1168.0 & $17.62\pm 0.19$ & $0.422 \pm 0.072$ & \\
&& 1177.8 &  1237.8 & $17.53\pm 0.16$ & $0.458 \pm 0.066$ & \\
&& 1246.9 &  1306.9 & $17.62\pm 0.22$ & $0.423 \pm 0.086$ & \\
&& 1316.3 &  1376.3 & $17.51\pm 0.17$ & $0.467 \pm 0.071$ & \\
&& 1386.0 &  1446.0 & $17.94\pm 0.24$ & $0.315 \pm 0.070$ & \\
&& 1455.2 &  1515.2 & $17.42\pm 0.16$ & $0.507 \pm 0.075$ & \\
&& 1524.6 &  1584.6 & $17.58\pm 0.16$ & $0.438 \pm 0.064$ & \\
&& 1594.3 &  1654.3 & $17.47\pm 0.11$ & $0.486 \pm 0.047$ & \\
&& 1663.9 &  1723.9 & $17.35\pm 0.13$ & $0.542 \pm 0.066$ & \\
&& 1733.0 &  1793.0 & $17.21\pm 0.11$ & $0.614 \pm 0.065$ & \\
&& 1802.6 &  2487.8 & $17.50\pm 0.07$ & $0.472 \pm 0.031$ & \\
&& 2496.9 &  3181.5 & $17.61\pm 0.06$ & $0.427 \pm 0.025$ & \\
&& 3190.8 &  3875.9 & $17.65\pm 0.06$ & $0.412 \pm 0.023$ & \\
&& 4180.7 &  4873.5 & $17.72\pm 0.06$ & $0.385 \pm 0.020$ & $0.000670 \pm 0.000080$\\
&& 4882.8 &  5567.3 & $17.90\pm 0.08$ & $0.325 \pm 0.023$ & $0.000615 \pm 0.000076$\\
&& 5576.4 &  6261.2 & $18.09\pm 0.09$ & $0.275 \pm 0.024$ & $0.000412 \pm 0.000066$\\
&& 6270.4 &  6608.3 & $17.60\pm 0.16$ & $0.431 \pm 0.065$ & \\
061007 & IIIa &    27.2 &    32.2 & $13.69\pm 0.07$ & $10.8 \pm  0.7$ & \\
&&   41.0 &    46.0 & $10.14\pm 0.02$ & $284 \pm   4$ & \\
&&   55.4 &    60.4 & $ 9.57\pm 0.02$ & $479 \pm   8$ & \\
&&   77.8 &    82.8 & $ 9.74\pm 0.01$ & $410 \pm   5$ & \\
&&   92.0 &    97.0 & $ 9.52\pm 0.02$ & $502 \pm   7$ & $1.61 \pm 0.05$\\
&&  105.9 &   110.9 & $ 9.64\pm 0.01$ & $450 \pm   5$ & $1.28 \pm 0.04$\\
&&  120.4 &   125.4 & $ 9.82\pm 0.03$ & $381 \pm   8$ & $1.03 \pm 0.03$\\
&&  134.6 &   139.6 & $ 9.97\pm 0.02$ & $332 \pm   4$ & $0.834 \pm 0.029$\\
&&  149.1 &   154.1 & $10.15\pm 0.02$ & $282 \pm   4$ & $0.736 \pm 0.027$\\
&&  163.8 &   168.8 & $10.36\pm 0.02$ & $232 \pm   4$ & $0.603 \pm 0.024$\\
&&  178.1 &   198.1 & $10.62\pm 0.02$ & $182 \pm   3$ & $0.422 \pm 0.012$\\
&&  207.4 &   227.4 & $10.84\pm 0.02$ & $149 \pm   2$ & $0.339 \pm 0.010$\\
&&  236.8 &   256.8 & $11.02\pm 0.01$ & $127 \pm   1$ & $0.272 \pm 0.009$\\
&&  266.0 &   286.0 & $11.21\pm 0.02$ & $106 \pm   1$ & $0.248 \pm 0.008$\\
&&  294.9 &   314.9 & $11.31\pm 0.02$ & $96.7 \pm  1.6$ & $0.210 \pm 0.007$\\
&&  332.0 &   352.0 & $11.56\pm 0.02$ & $77.3 \pm  1.3$ & $0.173 \pm 0.006$\\
&&  361.2 &   381.2 & $11.71\pm 0.02$ & $66.9 \pm  1.4$ & $0.162 \pm 0.006$\\
&&  390.3 &   410.3 & $11.85\pm 0.02$ & $59.2 \pm  1.3$ & $0.139 \pm 0.006$\\
&&  419.6 &   439.6 & $11.97\pm 0.02$ & $52.6 \pm  1.0$ & $0.126 \pm 0.005$\\
&&  456.5 &   476.5 & $12.16\pm 0.03$ & $44.3 \pm  1.1$ & $0.110 \pm 0.005$\\
&&  485.6 &   505.6 & $12.30\pm 0.03$ & $39.1 \pm  1.0$ & $0.0875 \pm 0.0042$\\
&&  514.5 &   534.5 & $12.40\pm 0.03$ & $35.7 \pm  0.9$ & $0.0880 \pm 0.0043$\\
&&  544.1 &   564.1 & $12.52\pm 0.03$ & $31.9 \pm  0.8$ & $0.0729 \pm 0.0038$\\
&& 1105.1 &  1110.1 & $13.79\pm 0.04$ & $9.91 \pm 0.35$ & $0.0238 \pm 0.0040$\\
&& 1119.0 &  1124.0 & $14.05\pm 0.12$ & $7.79 \pm 0.83$ & $0.0285 \pm 0.0044$\\
&& 1133.3 &  1138.3 & $13.94\pm 0.10$ & $8.59 \pm 0.80$ & $0.0238 \pm 0.0041$\\
&& 1147.8 &  1152.8 & $13.87\pm 0.09$ & $9.20 \pm 0.72$ & $0.0204 \pm 0.0037$\\
&& 1161.6 &  1166.6 & $14.15\pm 0.09$ & $7.08 \pm 0.58$ & $0.0299 \pm 0.0045$\\
&& 1175.8 &  1180.8 & $14.06\pm 0.09$ & $7.68 \pm 0.63$ & $0.0183 \pm 0.0035$\\
&& 1198.1 &  1203.1 & $13.94\pm 0.03$ & $8.59 \pm 0.26$ & $0.0217 \pm 0.0039$\\
&& 1212.2 &  1217.2 & $14.00\pm 0.09$ & $8.13 \pm 0.71$ & $0.0197 \pm 0.0037$\\
&& 1226.6 &  1231.6 & $14.07\pm 0.11$ & $7.62 \pm 0.75$ & $0.0258 \pm 0.0042$\\
&& 1241.2 &  1246.2 & $14.02\pm 0.09$ & $7.96 \pm 0.65$ & $0.0218 \pm 0.0040$\\
&& 1255.8 &  1275.8 & $14.13\pm 0.07$ & $7.23 \pm 0.46$ & $0.0230 \pm 0.0020$\\
&& 1285.3 &  1305.3 & $14.17\pm 0.09$ & $6.96 \pm 0.60$ & $0.0211 \pm 0.0019$\\
&& 1314.6 &  1334.6 & $14.22\pm 0.10$ & $6.66 \pm 0.59$ & $0.0199 \pm 0.0019$\\
&& 1344.0 &  1364.0 & $14.18\pm 0.06$ & $6.87 \pm 0.39$ & $0.0172 \pm 0.0018$\\
&& 1373.3 &  1393.3 & $14.29\pm 0.05$ & $6.25 \pm 0.31$ & $0.0190 \pm 0.0019$\\
&& 1402.4 &  1422.4 & $14.17\pm 0.06$ & $6.97 \pm 0.36$ & $0.0195 \pm 0.0019$\\
&& 1431.3 &  1451.3 & $14.38\pm 0.06$ & $5.74 \pm 0.31$ & $0.0179 \pm 0.0018$\\
&& 1460.2 &  1480.2 & $14.17\pm 0.09$ & $6.93 \pm 0.57$ & $0.0156 \pm 0.0017$\\
&& 1489.4 &  1509.4 & $14.33\pm 0.08$ & $6.01 \pm 0.42$ & $0.0170 \pm 0.0018$\\
&& 1518.5 &  1538.5 & $14.34\pm 0.06$ & $5.93 \pm 0.33$ & $0.0150 \pm 0.0017$\\
&& 1547.7 &  1567.7 & $14.40\pm 0.07$ & $5.62 \pm 0.36$ & $0.0167 \pm 0.0018$\\
&& 1576.8 &  1596.8 & $14.51\pm 0.06$ & $5.10 \pm 0.28$ & $0.0136 \pm 0.0016$\\
&& 1606.0 &  1888.7 & $14.68\pm 0.03$ & $4.34 \pm 0.13$ & $0.0128 \pm 0.0005$\\
&& 1897.9 &  2180.6 & $14.91\pm 0.02$ & $3.52 \pm 0.08$ & $0.00955 \pm 0.00050$\\
&& 2190.0 &  2471.1 & $15.14\pm 0.03$ & $2.85 \pm 0.09$ & \\
&& 2480.1 &  2763.3 & $15.33\pm 0.06$ & $2.39 \pm 0.13$ & \\
&& 2772.5 &  3135.5 & $15.57\pm 0.06$ & $1.92 \pm 0.10$ & \\
&& 3144.6 &  3426.8 & $15.67\pm 0.07$ & $1.75 \pm 0.11$ & \\
&& 3436.0 &  3718.2 & $15.78\pm 0.08$ & $1.57 \pm 0.12$ & \\
&& 3727.0 &  4018.1 & $15.82\pm 0.08$ & $1.52 \pm 0.12$ & \\
&& 4027.4 &  4310.4 & $16.02\pm 0.05$ & $1.27 \pm 0.06$ & \\
&& 4319.6 &  4602.3 & $16.11\pm 0.08$ & $1.17 \pm 0.09$ & \\
&& 4611.6 &  4893.8 & $16.12\pm 0.07$ & $1.15 \pm 0.08$ & \\
&& 4902.5 &  5192.9 & $16.35\pm 0.14$ & $0.938 \pm 0.124$ & \\
&& 5201.9 &  5485.3 & $16.05\pm 0.10$ & $1.23 \pm 0.12$ & \\
&& 5494.5 &  5776.0 & $16.46\pm 0.12$ & $0.845 \pm 0.092$ & $0.00128 \pm 0.00016$\\
&& 5785.1 &  6067.4 & $16.72\pm 0.10$ & $0.662 \pm 0.059$ & $0.00138 \pm 0.00016$\\
&& 6076.4 &  6358.3 & $16.97\pm 0.19$ & $0.529 \pm 0.095$ & $0.00103 \pm 0.00014$\\
&& 6367.2 &  6650.0 & $17.34\pm 0.20$ & $0.376 \pm 0.068$ & $0.000972 \pm 0.000130$\\
&& 6659.7 &  6941.5 & $16.84\pm 0.11$ & $0.596 \pm 0.059$ & $0.000860 \pm 0.000120$\\
&& 6950.5 &  7231.4 & $16.98\pm 0.16$ & $0.524 \pm 0.078$ & $0.000990 \pm 0.000131$\\
&& 7240.7 &  7522.4 & $16.77\pm 0.15$ & $0.636 \pm 0.087$ & $0.000837 \pm 0.000118$\\
&& 7531.7 &  7814.9 & $17.19\pm 0.22$ & $0.432 \pm 0.086$ & $0.000717 \pm 0.000107$\\
&& 7824.3 &  8106.9 & $17.32\pm 0.23$ & $0.382 \pm 0.082$ & $0.000827 \pm 0.000133$\\
&& 8116.3 &  8415.0 & $17.14\pm 0.23$ & $0.453 \pm 0.098$ & \\
&& 8424.2 &  8706.8 & $16.85\pm 0.14$ & $0.589 \pm 0.075$ & \\
&& 8716.0 &  8999.3 & $17.10\pm 0.18$ & $0.470 \pm 0.077$ & \\
&& 9008.1 &  9289.5 & $17.50\pm 0.22$ & $0.325 \pm 0.067$ & \\
&& 9298.7 &  9581.1 & $17.36\pm 0.25$ & $0.370 \pm 0.085$ & \\
&& 9590.3 &  9872.8 & $17.31\pm 0.17$ & $0.388 \pm 0.062$ & \\
&& 9882.2 & 10165.2 & $17.44\pm 0.17$ & $0.342 \pm 0.053$ & \\
&&10174.4 & 10464.9 & $17.47\pm 0.28$ & $0.332 \pm 0.087$ & \\
&&10473.9 & 10756.0 & $17.38\pm 0.30$ & $0.362 \pm 0.099$ & \\
&&10765.2 & 11047.0 & $17.76\pm 0.36$ & $0.255 \pm 0.083$ & \\
&&11055.8 & 11338.0 & $17.46\pm 0.18$ & $0.335 \pm 0.055$ & \\
&&11346.7 & 11629.0 & $> 17.56$ & $>0.306$ & \\
&&11637.8 & 11921.4 & $18.25\pm 0.47$ & $0.162 \pm 0.071$ & $0.000459 \pm 0.000072$\\
&&11930.6 & 12213.0 & $17.70\pm 0.32$ & $0.270 \pm 0.079$ & $0.000370 \pm 0.000063$\\
&&12221.8 & 12504.4 & $17.00\pm 0.14$ & $0.511 \pm 0.065$ & $0.000453 \pm 0.000071$\\
&&12513.6 & 12796.0 & $> 17.53$ & $>0.315$ & \\
&&12805.1 & 13087.3 & $17.92\pm 0.44$ & $0.220 \pm 0.090$ & $0.000395 \pm 0.000066$\\
&&13271.3 & 14137.1 & $17.57\pm 0.16$ & $0.303 \pm 0.045$ & $0.000343 \pm 0.000048$\\
&&14146.4 & 15051.9 & $17.80\pm 0.19$ & $0.246 \pm 0.043$ & \\
070611 & IIIc &    44.7 &   122.7 & $> 17.62$ & $>0.284$ & \\
&&  130.5 &   417.2 & $18.91\pm 0.37$ & $0.0872 \pm 0.0295$ & \\
&&  426.6 &  1111.3 & $20.05\pm 0.35$ & $0.0303 \pm 0.0098$ & \\
&& 1121.0 &  1805.7 & $19.42\pm 0.19$ & $0.0542 \pm 0.0093$ & \\
&& 1815.1 &  2152.7 & $18.47\pm 0.16$ & $0.131 \pm 0.019$ & \\
&& 2162.5 &  2499.9 & $18.22\pm 0.17$ & $0.164 \pm 0.025$ & \\
&& 2509.6 &  2847.5 & $18.50\pm 0.25$ & $0.127 \pm 0.029$ & \\
&& 2857.4 &  3195.2 & $18.54\pm 0.24$ & $0.123 \pm 0.027$ & \\
&& 3205.0 &  3889.9 & $18.68\pm 0.15$ & $0.108 \pm 0.015$ & $0.000232 \pm 0.000098$\\
&& 4202.6 &  4887.6 & $18.66\pm 0.11$ & $0.109 \pm 0.011$ & $0.0000965 \pm 0.0000441$\\
&& 4897.1 &  5581.7 & $> 19.39$ & $>0.0558$ & \\
&& 5591.1 &  6277.4 & $18.94\pm 0.22$ & $0.0846 \pm 0.0169$ & \\
&& 6286.7 &  6972.3 & $19.32\pm 0.27$ & $0.0599 \pm 0.0149$ & \\
&& 7829.1 &  8514.6 & $19.04\pm 0.17$ & $0.0769 \pm 0.0123$ & \\
&& 8524.1 &  9209.8 & $19.12\pm 0.30$ & $0.0714 \pm 0.0194$ & \\
&& 9219.6 &  9904.5 & $> 18.68$ & $>0.107$ & \\
\enddata
\tablecomments{Magnitudes are not corrected for Galactic extinction.  Optical
  flux densities ($\fnuo$ at 1.93 eV) have been corrected for Galacitc
  extinction and Ly$\alpha$ absorption in the IGM.  X-ray flux densities
  ($\fnux$ at 2.77 keV) are corrected for Galactic and host absorption.  All
  times are relative to $t_0$ given in \S~\ref{sec:observations}.}
\end{deluxetable}

\end{document}